\documentclass[twocolumn,showpacs,amsmath,amssymb]{revtex4}

\usepackage{graphicx}
\usepackage{dcolumn}
\usepackage{bm}

\begin{document}


\title{Cubic and non-cubic multiple-$q$ states in the Heisenberg
antiferromagnet on the pyrochlore lattice}

\author{Tsuyoshi Okubo}\email{okubo@spin.ess.sci.osaka-u.ac.jp}
\author{Trung Hai Nguyen }%
\author{Hikaru Kawamura}%
\affiliation{Department of Earth and Space Science,
 Faculty of Science, Osaka University, Toyonaka, Osaka 560-0043, Japan%
}%
\date{\today}

\begin{abstract}
The ordering of the classical Heisenberg model on the pyrochlore lattice  with the antiferromagnetic nearest-neighbor interaction $J_1$ and the  ferromagnetic next-nearest-neighbour interaction $J_2$ is investigated by means of a mean-field analysis and a Monte Carlo simulation. For a moderate  $J_2/J_1$-value, the model exhibits a first-order transition into an incommensurate  multiple-$q$  ordered state where multiple Bragg peaks coexist in the spin structure factor. We show that there are two types of metastable multiple-$q$ states, a cubic symmetric sextuple-$q$ state and a non-cubic symmetric quadruple-$q$ state. Based on a Monte Carlo simulation, we find that the cubic sextuple-$q$ state appears just below the first-order transition temperature, while another transition from the cubic sextuple-$q$ state to the non-cubic quadruple-$q$ state occurs at a lower temperature.
\end{abstract}

\pacs{75.10.Hk, 75.30.Kz, 75.50.Ee, 75.40.Mg}
\maketitle

\section{\label{sec:Intro} Introduction}

\begin{figure}
 \includegraphics[width=6cm]{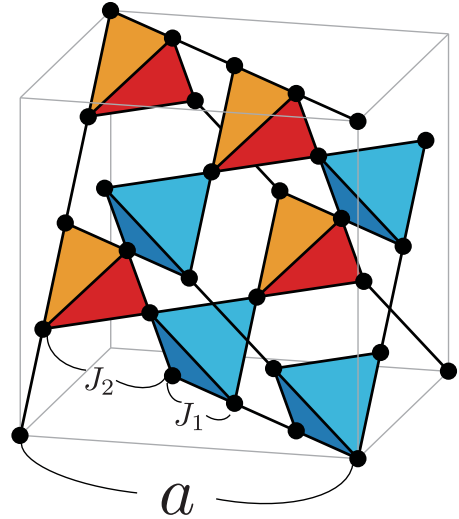}
 \caption{(color online) A pyrochlore lattice. The  nearest-neighbor interaction $J_1$ and the next-nearest-neighbor interaction  $J_2$ are indicated.}
 \label{fig_pyro}
\end{figure}

Recently, geometrically frustrated magnets have attracted much interest due to their unconventional ordering behaviors\cite{Book1,Book2,Journal_topic}. Spin systems on the pyrochlore lattice, which consists of a three-dimensional network of corner-sharing tetrahedra (Fig.\ref{fig_pyro}), are typical examples of such geometrically frustrated magnets. The classical Heisenberg magnet with the antiferromagnetic nearest-neighbor (NN) interaction is known to exhibit no magnetic long-range order even at zero temperature \cite{Reimers1992, Henley2005, Isakov2004, Conlon2009}. This is due to the macroscopic degeneracy of the ground state induced by geometrical frustration. Since such a high degeneracy is realized via a fine balance among frustrated interactions, it might be lifted by introducing small perturbations, {\it e.g.} the further-neighbor interactions\cite{Reimers1991,Tsuneishi,Nakamura,Chern,Conlon2010}, the quenched randomness\cite{Kastella, Saunders, Andreanov,Tam,Shinaoka}, or the lattice distortion\cite{Penc, Bergman, Channon}. The lifting of the degeneracy as a result of small perturbations might lead to an exotic magnetic state peculiar to geometrical frustration.

In this paper, we focus on the effects of the next-nearest neighbor (NNN) interaction on the pyrochlore-lattice classical Heisenberg model. Effects of the further neighbor interactions were investigated within a mean-field approximation by Reimers {\it et al.} up to the fourth neighbors \cite{Reimers1991}. In the case of the NNN interaction $J_2$ only, they showed that a $\bm{q}=0$ order was stabilized for the antiferromagnetic NNN interaction $J_2 < 0$, while an incommensurate $q$ order was stabilized for the ferromagnetic NNN $J_2 >0$. Since then, several Monte Carlo (MC) simulations on this $J_1$-$J_2$ pyrochlore Heisenberg model were performed \cite{Tsuneishi, Nakamura, Chern}. 

For the antiferromagnetic $J_2 < 0$, these works revealed that the system exhibited a first-order transition from the paramagnetic phase to the $\bm{q}=0$ ordered phase with a collinear up-up-down-down spin structure \cite{Tsuneishi,Chern}. Selection of such a collinear structure among possible $\bm{q}=0$ states might be due to ``order-by-disorder'' mechanism, where thermal fluctuations lift the degeneracy through an entropic contribution to the free energy \cite{Villain}.

For the ferromagnetic $J_2 > 0$, on the other hand, the situation seemed
to be more subtle. Tsuneishi {\it et al.} observed that the model with
$J_2/J_1=-0.1$ exhibited a first-order transition into a peculiar
ordered state, where enhanced spin fluctuations apparently coexisted with sharp
Bragg peaks associated with a multiple-$q$ structure \cite{Tsuneishi}. Chern {\it et al.} studied
the case of smaller $J_2/|J_1| \lesssim 0.09$ and showed that there occurred successive phase transitions with an intermediate phase characterized by a finite nematic order parameter and a layered structure, which was not predicted in the mean-field analysis \cite{Chern}.  Chern {\it et al.}  also pointed out that the low-temperature phase was a multiple-$q$ state which were basically the same as the one observed by Tsuneishi {\it et al}. In any case,  the explicit spin configuration of the multiple-$q$ state was not identified so far.

In the present paper, we investigate the nature of the multiple-$q$ state observed in the $J_1$-$J_2$ pyrochlore-lattice Heisenberg antiferromagnet with the ferromagnetic NNN interaction on the basis of a mean-field analysis and an extensive MC simulation. Particular attention is paid to the explicit spin configuration of the multiple-$q$ state and the nature of spin fluctuations in such state. We see that there are mainly two stable multiple-$q$ structures, the sextuple-$q$ state and the quadruple-$q$ state, each of which is characterized by whether they keep the cubic lattice symmetry or not. A first-order transition observed in earlier studies corresponds to the transition from the paramagnetic state to the cubic-symmetric multiple-$q$ (sextuple-$q$) state. We predict that another phase transition might occur at a lower temperature from the cubic-symmetric multiple-$q$ state to the non-cubic  multiple-$q$ (quadruple-$q$) state. Reflecting their characteristic spin fluctuations, internal fields in such multiple-$q$ states exhibit a broad distribution.

The rest of the paper is organized as follows. In Sec. \ref{Model}, we describe our model and briefly review the previous results on the model. In Sec. \ref{Mean}, we present the result of our mean-field calculation determining the explicit spin configurations of the possible multiple-$q$ states. Our MC results are presented in Sec. \ref{MC}. The results are discussed in conjunction with the mean-field result, including the explicit spin configurations of the multiple-$q$ states. Finally, we summarize our results in Sec. \ref{conclusion}.

\section{\label{Model}Model}

The model considered is the classical Heisenberg antiferromagnet on the
pyrochlore lattice, whose Hamiltonian is given by
\begin{equation}
 \mathcal{H} = -J_1\sum_{\langle i,j\rangle}\bm{S}_i\cdot\bm{S}_j-J_2\sum_{\langle\langle i,j\rangle\rangle}\bm{S}_i\cdot\bm{S}_j,
\end{equation}
where the first and the second sums are taken over all NN and NNN pairs $J_1$ and $J_2$ (see Fig.\ref{fig_pyro}), respectively. We suppose the antiferromagnetic NN interaction $J_1 <0$ and the ferromagnetic NNN interaction $J_2 >0$ ($J_1$-$J_2$ model), and $a$ to be the length of the cubic unit cell. Note that the pyrochlore lattice, which can be viewed as an FCC lattice formed by regular tetrahedra of a fixed orientation, is a non-Bravais lattice.  

 The ordering of this $J_1$-$J_2$ model was first studied by Reimers {\it et al.} based on a mean-field approximation. There, the full density matrix of the system was approximated by a product of single spin density matrices with local effective fields which were determined so as to minimize the free energy. Reimers {\it et al.} found that the model exhibited a phase transition from the paramagnetic phase to a magnetically ordered phase, where the magnetic long range order was characterized by twelve incommensurate wavevectors of the FCC Bravais lattice: $(q^\ast, q^\ast, 0)$,$ (q^\ast, -q^\ast, 0)$, $ (0, q^\ast, q^\ast)$,$ (0, q^\ast, -q^\ast)$, $(q^\ast, 0, q^\ast)$,$ (-q^\ast, 0, q^\ast)$ and their minus, with $q^\ast \simeq \frac{3\pi}{2a}$ \cite{Reimers1991}. Since the pyrochlore lattice consists of four FCC sublattices, each wavevector has four independent eigenmodes and only a certain combination of them becomes unstable at the transition point.

 In fact, there are infinitely many ways to mix such twelve critical modes. Thus, just the determination of unstable modes is not enough to specify the explicit spin configuration of the ordered phase. Indeed, Reimers {\it et al.} did not determine the ordered spin configuration, nor specified even whether it was a single-$q$ state or a multiple-$q$ state.

Multiple-$q$ state is generally incompatible with the fixed spin-length condition  $|\bm{S}_i|=1$ so that it is not favored in the classical spin system  at low enough temperatures where the fixed spin-length condition needs to be observed. Meanwhile, the unstable critical mode of the present model entails the magnitudes of frozen spin moments differ from one sublattice to the other so that even a simple single-$q$ state is incompatible with the fixed spin-length condition. It means that the multiple-$q$ state might have a higher chance to be stabilized, particularly at moderate temperatures where thermal fluctuations play a role. 

 As mentioned, a multiple-$q$ ordered state characterized by multiple Bragg peaks in the spin structure factor was reported in previous MC simulations on the model \cite{Tsuneishi, Chern}. Interestingly, Tsuneishi {\it et al.} also reported that this multiple-$q$ ordered state accompanies only small amount of spin freezing, and some sort of spin fluctuations apparently coexisted with sharp Bragg peaks. The detailed spin structure of the multiple-$q$ state, however, has not been clarified so far. Especially, the origin of the spin fluctuation reported by Tsuneishi {\it et al.} remains unsolved.

Chern {\it et al.} reported that yet another phase, a partially ordered collinear phase, might be realized between the paramagnetic phase and the multiple-$q$ phase for sufficiently small values of $|J_2/J_1| \lesssim 0.09$ \cite{Chern}. This partially ordered collinear phase is characterized by a finite nematic order parameter and a layered spin structure. Note that this phase is different from the multiple-$q$ ordered phase discussed above where the system retains a magnetic long range-order characterized by sharp Bragg peaks. Although the partially ordered collinear phase as discussed by Chern {\it et al.} is also interesting, we focus in the present paper on the nature of the multiple-$q$ ordered state and of the associated fluctuations in the multiple-$q$ ordered phase.

\section{\label{Mean}Mean-field approximation}

In studying the nature of the multiple-$q$ ordered state, we begin by
revisiting a mean-field analysis done earlier by Reimers {\it et al.}
\cite{Reimers1991}. Reimers {\it et al.} constructed a Landau-type free
energy $F$ of the pyrochlore Heisenberg antiferromagnet within a
mean-field approximation up to the quartic order,
\begin{multline}
 F/N_s = -4T\ln 4\pi
 \\+\frac{1}{2}\sum_{\bm{q}}\sum_{\mu\nu}\bm{B}^{(\mu)}_{\bm{q}}\cdot\bm{B}^{(\nu)}_{-\bm{q}}(3T\delta^{\mu\nu}-J_{\bm{q}}^{\mu\nu})
\\
 +\frac{9T}{20}\sum_\mu\sum_{\left\{\bm{q}\right\}}\hspace{-0.45em}\raisebox{0.4em}{$~^\prime$}\left(\bm{B}_{\bm{q}_1}^{(\mu)}\cdot\bm{B}_{\bm{q}_2}^{(\mu)}\right) \left(\bm{B}_{\bm{q}_3}^{(\mu)}\cdot\bm{B}_{\bm{q}_4}^{(\mu)}\right),
\end{multline}
where $\bm{B}_{\bm{q}}^{(\mu)}$ is the order parameter corresponding to the
 Fourier magnetization of sublattice $\mu$ ($\mu=1,2,3,4$) given by
\begin{align}
 \bm{B}_{\bm{q}}^{(\mu)} &= \langle \bm{S}_{\bm{q}}^{(\mu)}\rangle,\notag \\
 \bm{S}_{\bm{q}}^{(\mu)} &= \frac{1}{N_s}\sum_{i}\bm{S}_{i}^{(\mu)}
 \exp(-\bm{q}\cdot\bm{r}_i^{(\mu)}),
\end{align}
where $\bm{S}_{i}^{(\mu)}$ is the spin at the site $i$ belonging to
sublattice $\mu$, $\bm{r}_{i}^{(\mu)}$ is the position vector of that
site and $N_s$ is the number of spins belonging to sublattice $\mu$. The sum $\sum_{\left\{\bm{q}\right\}}'$ runs over all $\bm{q}_i$s which  satisfy $\bm{q}_1+\bm{q}_2+\bm{q}_3+\bm{q}_4=\bm{0}$, and $J_{\bm{q}}^{\mu\nu}$ is  the Fourier transform of the exchange interaction between sublattices  $\mu$ and $\nu$. The quadratic term can be diagonalized by a unitary matrix $U_{\bm{q}}$ with the eigenvalue $\lambda_{\bm{q}}$,
\begin{equation}
 \sum_\nu
  J_{\bm{q}}^{\mu\nu}U_{\bm{q}}^{\nu i}=\lambda_{\bm{q}}^iU_{\bm{q}}^{\mu i}.
\end{equation}
where $i$ indicates each eigenmode. By transforming the order parameter to normal modes $\bm{\Phi}_{\bm{q}}^i$,
\begin{equation}
 \bm{B}_{\bm{q}}^{(\mu)}=\sum_{i} U_{\bm{q}}^{\mu i}\bm{\Phi}_{\bm{q}}^i,
\end{equation}
the Landau free energy $F$ is reduced to
\begin{multline}
 F/N_s = -4T\ln 4\pi
 \\+\frac{1}{2}\sum_{\bm{q}}\sum_{i}\left|\bm{\Phi}^i_{\bm{q}}\right|^2(3T-\lambda_{\bm{q}}^{i})
\\
 +\frac{9T}{20}\sum_{ijkl}\sum_{\left\{\bm{q}\right\}}\hspace{-0.45em}\raisebox{0.4em}{$~^\prime$}\left(\bm{\Phi}_{\bm{q}_1}^i\cdot\bm{\Phi}_{\bm{q}_2}^j\right)
 \left(\bm{\Phi}_{\bm{q}_3}^k\cdot\bm{\Phi}_{\bm{q}_4}^l\right) \\
 \times
 \sum_{\mu}U_{\bm{q_1}}^{\mu i}U_{\bm{q_2}}^{\mu j}U_{\bm{q_3}}^{\mu
 k}U_{\bm{q_4}}^{\mu l}.
 \label{Landau}
\end{multline}

 From the quadratic term of the free-energy expansion, one sees that the
 normal mode corresponding to the maximum eigenvalue
 $\lambda_{\bm{q}^\ast}^i$ becomes unstable at $T_c = \frac{1}{3} \lambda_{\bm{q}^\ast}^i$ where $\bm{q}^\ast$ is  the critical wavevector, leading to a phase transition to the ordered state characterized by the wavevector $\bm{q}^\ast$. When the maximum eigenvalue is degenerate as in the present case, the ordered-state spin configuration still remains largely undetermined at the quadratic level. In such a case, there  are infinitely many ways of mixing  $\bm{\Phi}_{\bm{q}^\ast}^i$s with keeping the order parameter
\begin{equation}
 m^2 \equiv  \sum |\bm{\Phi}^i_{\bm{q}^\ast}|^2
\end{equation}
constant. In order to specify the explicit ordered-state spin configuration, one needs  to go to the quartic term. Reimers {\it et al.\/} made such an analysis  only for the special case of a simple $\bm{q}=\bm{0}$ ordered state, whereas such  an analysis has not been made for more general cases of an incommensurate ordered state, which is the target of our following analysis.

 Now, we wish to go beyond the analysis of Reimers {\it et al.} to derive the explicit ordered-state spin configuration of the $J_1$-$J_2$ model within a mean-field approximation. For the case of $J_1 <0$ and $J_2 >0$ of our interest, the maximum eigenvalue of $J_{\bm{q}}$ appears in a symmetric direction $\bm{q}=(q,q,0)$ and its cubic-symmetry counterparts. Along this direction, $J_{\bm{q}}$ has a form
\begin{equation}
 J_{\bm{q}} = 2 \begin{pmatrix}
       0& J^{(1)}& J^{(2)}&J^{(2)}\\
       J^{(1)}& 0& J^{(2)}&J^{(2)}\\
       J^{(2)}&J^{(2)}&0& J^{(3)}\\
       J^{(2)}&J^{(2)}&J^{(3)}&0
      \end{pmatrix},
\end{equation}
with
\begin{align}
 J^{(1)} &= J_1\cos\frac{q}{2} + 2J_2, \notag \\
 J^{(2)} &= J_1\cos\frac{q}{4} +
       J_2\left(\cos\frac{q}{4} +\cos{\frac{3}{4}q} \right), \notag \\
 J^{(3)} &= J_1 + 2J_2\cos\frac{q}{2}.
\end{align}
The eigenvalues of this matrix are calculated as
\begin{widetext}
\begin{align}
 \lambda_1&= -2J_1\cos\frac{q}{2}-4J_2,\\
 \lambda_2&= - 2J_1-4J_2\cos\frac{q}{2},\\
\lambda_\pm&= 
 \left(J_1+2J_2\right)\left(\cos\frac{q}{2}+1\right) \pm
 \sqrt{\left(J_1-2J_2\right)^2\left(\cos\frac{q}{2}-1\right)^2+16\left[J_1\cos\frac{q}{4}+J_2\left(\cos\frac{q}{4}+\cos\frac{3q}{4}\right)\right]^2},
\end{align}
 \end{widetext}
with the corresponding eigenvectors given by
\begin{equation}
 \vec{U}^1_{q} = \frac{1}{\sqrt{2}}\begin{pmatrix}
                                1\\
                                -1 \\
                               0\\
                               0\\
                               \end{pmatrix},
 \vec{U}^2_{q} = \frac{1}{\sqrt{2}}\begin{pmatrix}
                                0\\
                                0 \\
                               1\\
                               -1\\
                               \end{pmatrix},
\end{equation}
and
\begin{equation}
 \vec{U}^\pm_{q} = \frac{1}{\sqrt{2+2\alpha^2_\pm}} \begin{pmatrix}
                                            1\\
                                            1\\
                                            -\alpha_\pm\\
                                            -\alpha_\pm\\
                               \end{pmatrix},
\label{eq_eigen_vec}
\end{equation}
with 
\begin{equation}
 \alpha_\pm(q) = \frac{-\lambda_\pm +2\left(J_1\cos\frac{q}{2}+2J_2\right)}{4\left[J_1\cos\frac{q}{4}+J_2\left(\cos\frac{q}{4}+\cos\frac{3q}{4}\right)\right]}.
\label{alpha}
\end{equation}

The maximum engenvalue lies in the $\lambda_{+}$ branch at a wavevector $q=q^\ast\simeq \frac{3\pi}{2a}$. Note that, since $|\alpha_{+}|\neq 1$ generally in the corresponding eigenvector $\vec{U}^{+}_{q}$, a single-$q$ state associated with such an eigenvector cannot satisfy the fixed spin-length condition $|\bm{S}_i|=1$ for all sublattices. Hence, even a pure single-$q$ state cannot be realized at  $T= 0$ as in multiple-$q$ states, though it might be realized at finite temperatures due to thermal fluctuations.

Due to the cubic symmetry of the pyrochlore lattice, $\lambda_+(q^\ast)$ is twelvefold degenerate. Hence, a variety of multiple-$q$ states might be possible in principle. Even if one is to identify two eigenmodes corresponding to $\bm{q}$ and $-\bm{q}$, there are still six independent eigenmodes. Within the Landau-type free-energy expansion \eqref{Landau}, the free-energy difference among possible multiple-$q$ states arises from the quartic term,
\begin{multline}
 f_4 =  \frac{9T}{20}\sum_{i,j,k,l}\sum_{\left\{\bm{q}\right\}}\hspace{-0.45em}\raisebox{0.4em}{$~^\prime$}\left(\bm{\Phi}_{\bm{q}_1}^i\cdot\bm{\Phi}_{\bm{q}_2}^j\right)  \left(\bm{\Phi}_{\bm{q}_3}^k\cdot\bm{\Phi}_{\bm{q}_4}^l\right) \\
 \times
 \sum_{\mu}U_{\bm{q_1}}^{\mu i}U_{\bm{q_2}}^{\mu j}U_{\bm{q_3}}^{\mu
 k}U_{\bm{q_4}}^{\mu l}.
 \label{eq_f4}
\end{multline}
where the sum over $i,j,k,l$ is taken over eigenmodes ($i,j,k,l=1,2,\pm$). Just below the transition temperature, one might neglect in the summation $\sum_{ijkl}\sum_{\bm\left\{\bm{q}\right\}}'$ the contributions from all modes other than the twelve critical modes. Let us represent independent critical wavevectors as
\begin{align}
 \bm{q}_1^{\pm}&= (q^\ast,\pm q^\ast,0) \notag \\
 \bm{q}_2^{\pm}&= (0,q^\ast,\pm q^\ast) \notag \\
 \bm{q}_3^{\pm}&= (\pm q^\ast,0,q^\ast).
 \label{eq_qis}
\end{align}
When one gives the number of critical modes to be mixed, there remain degrees of freedom associated with the amplitudes and the directions of the critical modes $\bm{\Phi}_{\bm{q}_i^{\pm}}^+$, which should be determined so as to minimize $f_4$.
\begin{figure*}
 \includegraphics[width=16cm]{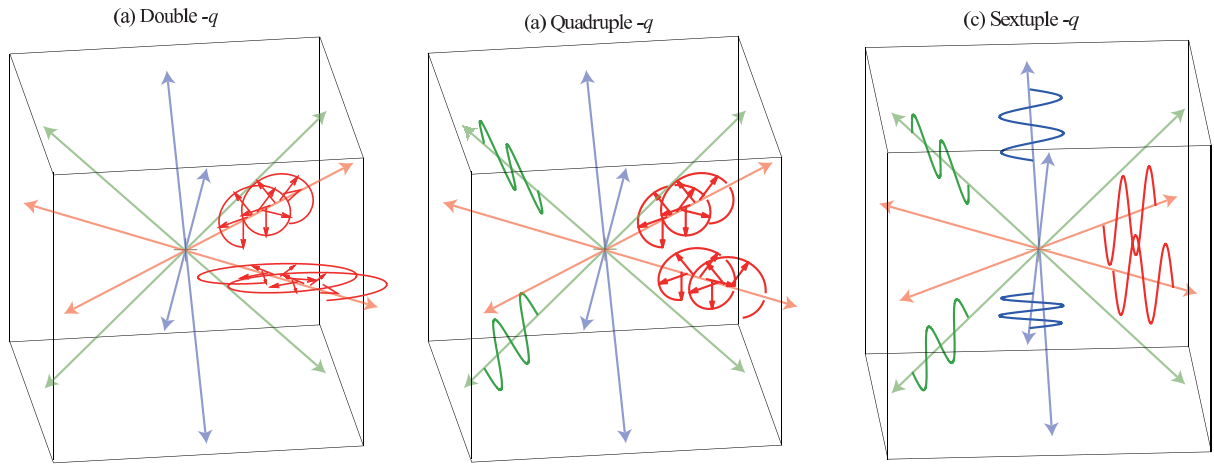}
 \caption{(color online) Schematic representation of several multiple-$q$ states. Twelve arrows represent twelve wavevectors corresponding to critical eigenmodes of the model. A linearly-polarized spin-density wave (SDW) or a spiral depicted on each arrow demonstrates the type of the corresponding eigenmode. (a) The double-$q$ state which is a superposition of two spirals of $\bm{q}_i^{\pm}$ with mutually orthogonal spiral planes. (b) The quadruple-$q$ state which is a superposition of two spirals of $\bm{q}_i^{\pm}$ sharing a common spiral plane and two SDWs perpendicular to the spiral plane. (c) The sextuple-$q$ state which is a superposition of six SDWs where each pair of $\bm{q}_i^\pm$ shares a common axis, while axes of different $\bm{q}_i$s are mutually orthogonal.}
\label{fig_multiple}
\end{figure*}

 In case of the single-$q$ state, $f_4$ takes a minimum for a simple spiral state
\begin{equation}
 \bm{S}(\bm{r}) \propto \cos(\bm{q}^\sigma_{i}\cdot\bm{r} +
                        \theta)\bm{e}_1 \pm
                        \sin(\bm{q}^\sigma_{i}\cdot\bm{r}+\theta)\bm{e}_2,
\end{equation}
where $\sigma=\pm$ with $\bm{e}_1\perp\bm{e}_2$ and $|\bm{e}_1|=|\bm{e}_2|=1$, while the minimized $f_4$ is given by
\begin{equation}
 f^{(\mathrm{si})}_4= \frac{9T}{10(1+\alpha^2)^2}
 \left(1+\alpha^4\right)m^4.
\end{equation}
where $\alpha\equiv \alpha_+(q^\ast)$.
 
 In addition to the simple single-$q$ state, various types of multiple-$q$ states are possible depending on the  number of mixed critical modes. The multiple-$q$ states with odd number of  critical modes, {\it i.e.\/}, the states with three or five  mixed eigenmodes, turn out to have higher $f_4$, and we consider here
 only the multiple-$q$ states with even number of critical modes, {\it
 i.e.\/}, the states with two, four and six mixed eigenmodes, which are
 described as the double-$q$, the quadruple-$q$, and the sextuple-$q$
 states, respectively. For a given even number of mixed critical modes, one
 needs to minimize $f_4$ for a fixed order parameter $m^2$ with respect
 to the amplitudes and the directions of the critical modes
 $\bm{\Phi}_{\bm{q}_i^{\pm}}^+$. Some of the details of this
 minimization procedure are given in appendix.  

In case of the double-$q$ state, the optimized spin configuration turns
out to be a superposition of two {\it distorted\/} spirals of $\bm{q}_i^{\pm}$ which have mutually orthogonal spiral planes as illustrated in Fig.\ref{fig_multiple} (a) (the d1 state in appendix). The minimized  $f_4$ is calculated to be
\begin{equation}
 f_4^{(\mathrm{d})}= \frac{9T}{40\left(1+\alpha^2\right)^2}
  \left[3+4\alpha^2+3\alpha^4 -\frac{\left(1+\alpha^4\right)^2}{\left(1+\alpha^2\right)^2} \right]m^4.
\end{equation}

In case of the quadruple-$q$ state, the optimized spin configuration turns out to be a superposition of two spirals $\bm{q}_i^{\pm}$ sharing a common spiral plane and two linearly-polarized spin-density waves (SDW), $\bm{S}(\bm{r}) \propto \cos(\bm{q}_j^{\sigma}\cdot\bm{r} + \theta)\bm{e}_3$, with its spin polarization perpendicular to the spiral plane as illustrated in Fig.\ref{fig_multiple} (b). The minimized $f_4$ is calculated to be
\begin{multline}
 f_4^{(\mathrm{q})} =
 \frac{9T}{40\left(1+\alpha^2\right)^2}\Biggl[-\frac{\left(1+12\alpha^2+\alpha^4\right)^2}{1+20\alpha^2+\alpha^4}\\
                                       +3\left(1+4\alpha^2+\alpha^4\right)\Biggr]m^4.
\end{multline}

Finally, in case of the sextuple-$q$ state, the optimized spin configuration turns out to be a superposition of six linearly-polarized SDWs where each pair of $\bm{q}_i^\pm$ shares a common axis $\bm{e}_i$, and axes of different $\bm{q}_i$s are mutually orthogonal $\bm{e}_1\perp \bm{e}_2\perp \bm{e}_3$ as illustrated in Fig.\ref{fig_multiple} (c). Note that this sextuple-$q$ state retains a cubic symmetry of the pyrochlore lattice, in sharp contrast to the other multiple-$q$ states or the single-$q$ state which break the cubic lattice symmetry. The minimized  $f_4$ is calculated to be
\begin{equation}
 f_4^{\mathrm{(se)}}=\frac{3T}{40(1+\alpha^2)^2}\left(7+12\alpha^2+7\alpha^4\right)m^4.
\end{equation}

\begin{figure}
 \includegraphics[width=8cm]{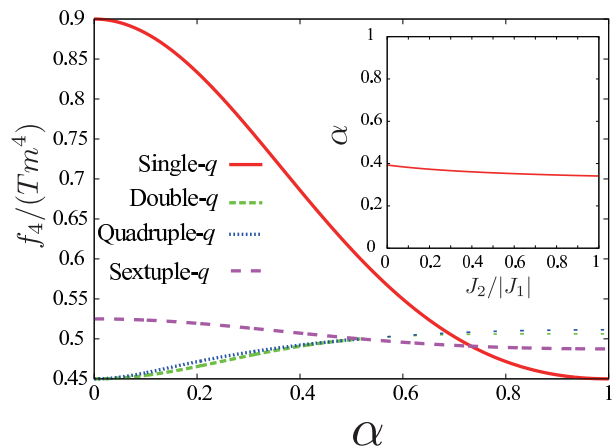}
 \caption{(color online) Quadratic terms of the free energy $f_4$ of the single-$q$  state and of various types of multiple-$q$ states, including the double-$q$ state (the $\mathrm{d1}$ state in appendix), the quadruple-$q$ state (the $\mathrm{q2}$ state in appendix) and the sextuple-$q$ state, are plotted versus
 the parameter $\alpha=\alpha^+(\bm{q}^\ast)$: See Eq.\eqref{alpha}. Inset exhibits a relation between $\alpha$ and
 $J_2/|J_1|$. For $\alpha \gtrsim 0.518$ the double-$q$ state and
the quadruple-$q$ state become locally unstable. Such unstable regions are indicated by
 thin dotted lines. For further details of the stability, see appendix.}
 \label{fig_Free_ene}
\end{figure}

In Fig.\ref{fig_Free_ene}, we show $f_4/T$ of various states discussed above, including the single-$q$ state, the double-$q$ state, the quadruple-$q$ state and the sextuple-$q$ state, as a function of the parameter $\alpha$. In fact, $\alpha$ depends on $J_2/J_1$ only weakly. For the value of $J_2 < |J_1|$, $\alpha$ takes values around $\alpha \simeq 0.4$: See the inset of Fig.\ref{fig_Free_ene}. Around this value, the free energy of the single-$q$ state turns out to be much higher than those of the multiple-$q$ states. This result is consistent with the observation of the multiple-$q$ states in previous MC simulations \cite{Tsuneishi,Chern}. Within the present mean-field approximation, the double-$q$ state becomes most stable for $\alpha \simeq 0.4$.  The free energy difference among various multiple-$q$ states, however, turns out to be rather small, and fluctuations neglected here might eventually select the multiple-$q$ state other than the double-$q$ state. In order to determine the true ordered state of the model, we should carefully examine the effect of fluctuations. For this purpose, we perform extensive MC simulations of the model in the next section.

\section{Monte Carlo simulation\label{MC}} 

In this section, we present the results of our MC simulation on the $J_1$-$J_2$ pyrochlore-lattice classical Heisenberg model with $J_1<0$ and $J_2>0$. As a typical example, we deal with the case of $J_2/J_1=-0.2$. In the phase diagram reported by Chern {\it et al.}\cite{Chern}, this $J_2/|J_1|$ ratio is sufficiently large so that the partially ordered collinear phase should not appear. A direct transition from the paramagnetic phase to the multiple-$q$ ordered phase is then expected.

The pyrochlore lattice contains 16 spins in its cubic unit cell. The system size we deal with is of linear size $L$ in units of cubic unit cell, {\it i.e.} the system contains $N=16L^3$ spins in total. Since the ordered state is generally  incommensurate with the underlying lattice, the system under periodic boundary conditions (BC) would be subject to severe finite size effects. In order to examine such finite-size effects, we also consider free BC, and carefully compare the results between these two BC. In the case of free BC, we extend the lattice with a half of the unit cell in all three directions so that it keeps the cubic symmetry of the lattice. Then, the system contains $N=2 ((2L+1)^3+1) $ spins in total.

MC simulations are performed based on the standard heat-bath method combined with the over-relaxation method. The lattice size is of $8 \le L \le 24$ for periodic BC, and of $8 \le L \le 32$ for free BC. The system is gradually cooled from the high temperature. A run at each temperature contains typically $(2-4) \times 10^5$ MC steps per spin (MCS) and about a half of MCS are discarded for equilibration. Our 1 MCS consists of 1 heat-bath sweep and subsequent 10 over-relaxation sweeps.

\begin{figure}
\includegraphics[width=7cm]{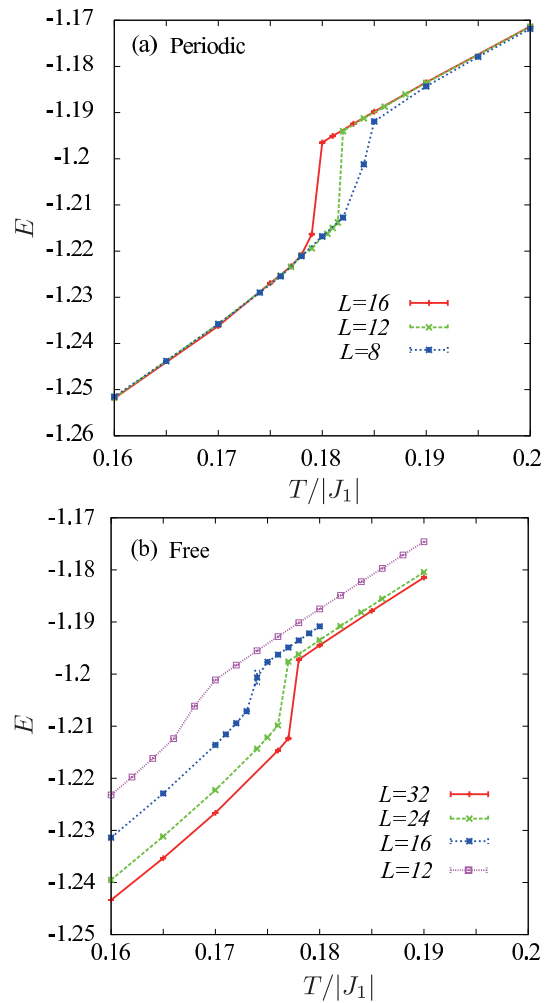}
\caption{(color online) Energy per spin versus the temperature for both cases of periodic  boundary conditions (a), and free boundary conditions (b). The interaction parameter is $J_2/J_1 = -0.2$.}
\label{fig_ene}
\end{figure}

In Fig.\ref{fig_ene}, we show the temperature and size dependence of the energy per spin. For both cases of periodic and free BC,  an almost discontinuous change of the energy, a characteristics of a first-order transition, is observed. Such a first-order nature of the transition is consistent with the results of previous calculations \cite {Tsuneishi, Chern}. We estimate the bulk transition temperature to be $T_c \simeq 0.178 |J_1|$.

At temperatures below $T_c$, several types of metastable states are observed in our simulations, depending on the spin initial conditions and the random-number sequences. In order to probe the spin configurations in these metastable states, we calculate the spin structure factor $F(\vec{q})$ defined by
\begin{equation}
 F(\bm{q}) \equiv \frac{1}{N}\left\langle \left|\sum_{i}\sum_{\mu} \bm{S}^{(\mu)}_i
 e^{i\bm{q}\cdot\bm{r}^{(\mu)}_i}\right|^2 \right\rangle,
\end{equation}
where $\langle \cdots \rangle $ represents a thermal average. From the spin structure factor, we find that none of metastable states is a single-$q$ state. All of them are multiple-$q$ states where the multiple magnetic Bragg peaks coexist in the spin structure factor.

\begin{figure*}
 \includegraphics[width=16cm]{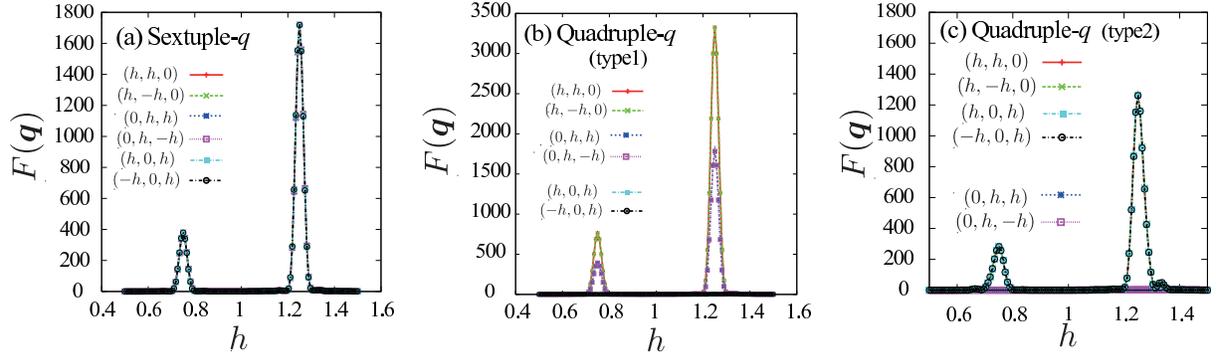}
\caption{(color online) The spin structure factor of the multiple-$q$ states along the  directions of $\bm{q}=\frac{2\pi}{a}(h, h, 0 )$ and of its
 cubic-symmetry counterparts, calculated at $T=0.17|J_1|$ with
 $J_2/J_1=-0.2$. The lattice is of $L=16$ with periodic boundary
 conditions. Each curve corresponds to different directions of
 wavevectors related via the cubic symmetry of the lattice. (a) The
 sextuple-$q$ state where Bragg peaks appear in all six directions with
 the same intensity. (b) The quadruple-$q$ state (of type 1) where Brag
 peaks appear only four out of six directions. Half of main Bragg peaks
 corresponding to the wavevectors in the same plane ($h,\pm h, 0$) are larger in magnitude than those in the other plane ($0,h,\pm h$). (c) The quadruple-$q$ state (of type 2) where Brag peaks appear in four out of six directions with the same intensity. In this type 2 structure, all Bragg peaks including main and submain are split into two, one at $(q, q, 0)$ and the other at $(q-\frac{2\pi}{L}, q, 0)$. The peaks shown in the figure are the one at $(q, q, 0)$.}
 \label{fig_Sq}
\end{figure*}

\begin{figure*}
 \includegraphics[width=12cm]{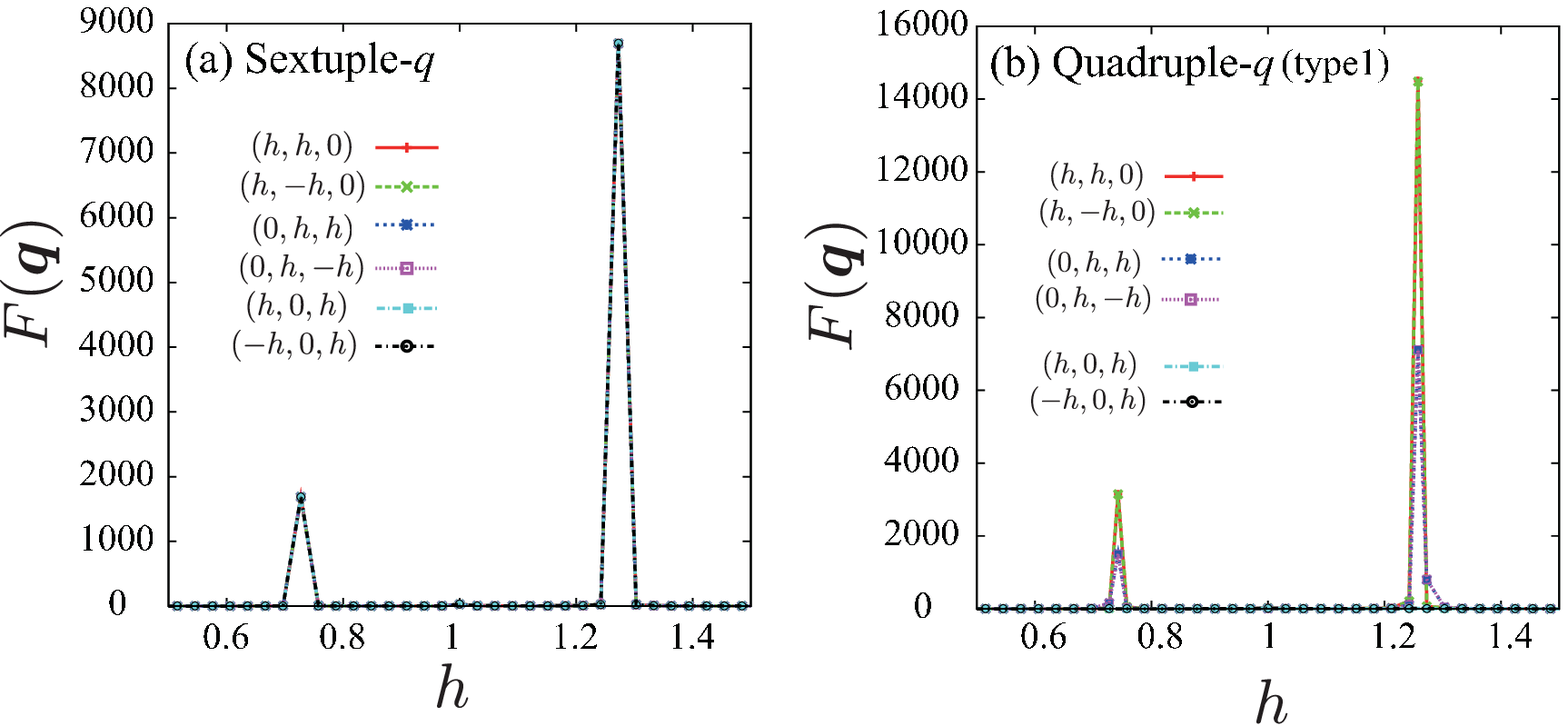}
\caption{(color online) The spin structure factor of the multiple-$q$ states along the
 directions of $\bm{q}=\frac{2\pi}{a}(h, h, 0 )$ and of its
 cubic-symmetry counterparts, calculated at $T=0.17|J_1|$ with
 $J_2/J_1=-0.2$. The lattice is of $L=32$ with free boundary
 conditions. Each curve corresponds to different directions of
 wavevectors related via the cubic symmetry of the lattice. (a) The
 sextuple-$q$ state where Bragg peaks appear in all six directions with
 the same intensity. (b) The quadruple-$q$ state (of type 1) where Brag
 peaks appear only four out of six directions. In (b), the main and sub
 peaks appear along the direction $\bm{q} = \frac{2\pi}{a}(h,
 h +\delta, 0)$, slightly off the high symmetric direction $(h,
 h, 0)$, where $\delta=1/(L+1)$ is simply the mesh size of our measurements in the wavevector space. The peak height shown here is of the one at $(h, h +\delta, 0)$, while the peak position is set at $h=\sqrt{\left[h+(h+\delta)^2\right]/2}$.
}
\label{fig_Sq_free}
\end{figure*}

In Fig.\ref{fig_Sq} we show the typical spin structure factors of these metastable states calculated at a temperature $T=0.17|J_1| < T_c\simeq 0.178|J_1|$ under periodic BC. These metastable states might be classified into two types according as whether they keep the cubic symmetry or not. In the cubic-symmetric state, there are six independent main Bragg peaks with the same intensity at $\bm{q} = \frac{2\pi}{a}(h^\ast, h^\ast, 0 )$ and at its cubic-symmetry counterparts with $h^\ast = 5/4$ (see Fig.\ref{fig_Sq} (a)). The peak position $h^\ast$ is close to the corresponding mean-field value $h^\ast \simeq 1.263$. There are also other Bragg peaks related to the main peaks via the reciprocal lattice vectors of the FCC lattice.

Since the observed cubic-symmetric state involves six critical wavevectors, we identify this state as a sextuple-$q$ state. In case of periodic BC, we sometimes observe ``almost cubic symmetric'' states where the position of the Bragg peak shifts by $\frac{2\pi}{L}$, or one of the Bragg peaks splits into two, due to the finite size effects associated with periodic BC. We regard them as modified forms of the sextuple-$q$ state.

In the observed non-cubic state, two out of six main Bragg peaks with
wavevectors on a common plane, {\it e.g.\/}, $(q,0,q)$ and $(-q,0,q)$,
vanish, while other four remain. Since there are four critical
wavevectors in this non-cubic state, we identify the state as a
quadruple-$q$ state. Two different types of the non-cubic states are
observed, depending on the intensity ratio among the remaining four main
Bragg peaks. In one type (type 1), two out of four remaining Bragg peaks with
wavevectors on a common plane, {\it e.g.\/},  $(q,q,0)$ and $(q,-q,0)$,
have stronger intensity than the other two with wavevectors lying in the
other plane, {\it e.g.\/}, $(0,q,q)$ and $(0,q,-q)$: See
Fig.\ref{fig_Sq} (b). In the other type (type 2), four main Bragg peaks have
equal intensities (see Fig.\ref{fig_Sq} (c)). In this type 2 structure, all Bragg peaks including main and submain are actually split into two, one at $(q, q, 0)$ and the other at $(q-\frac{2\pi}{L}, q, 0)$. 

\begin{figure}
\includegraphics[width=8cm]{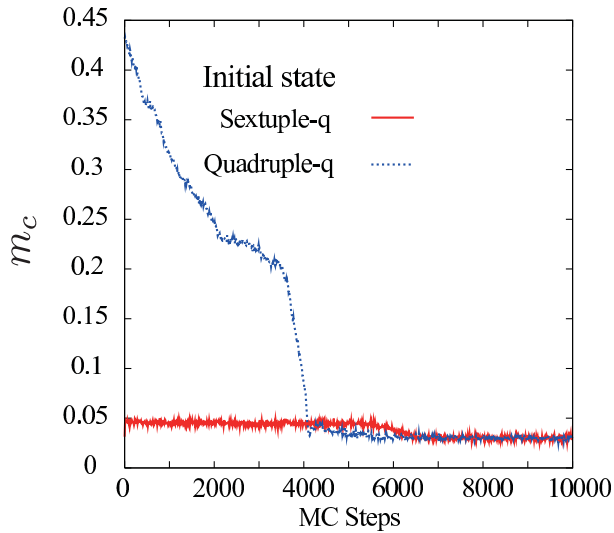}
\caption{(color online) The Monte-Carlo time evolution of the cubic-symmetry-breaking order parameter $m_c$ defined by Eq.\eqref{eq_mc}, calculated at $T=0.15|J_1|$ for a $L=48$ lattice with free boundary conditions. The interaction parameter is $J_2/J_1=-0.2$. Initially, a half side  of the system is prepared to be a sextuple-$q$ state, while the other half a quadruple-$q$ state. The two curves represent the subsequent time evolution of the cubic-symmetry-breaking order parameter $m_c$, each calculated at each half side of the system. The data indicates that the cubic state expands as the simulation goes on.}
\label{fig_mix}
\end{figure}

\begin{figure}
\includegraphics[width=8cm]{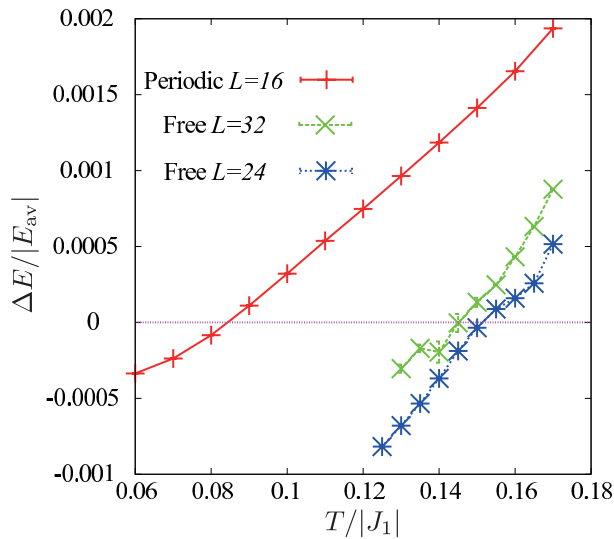}
\caption{(color online) Energy difference between the sextuple-$q$ state and the quadruple-$q$
 state plotted versus the temperature for both cases of periodic and
 free boundary conditions. The interaction parameter is
 $J_2/J_1=-0.2$. In case of periodic boundary conditions, we choose a
 pair of metastable states which have the lowest energy among several
 ``almost cubic-symmetric'' states or ``almost quadruple-$q$ (type 1)''
 states where peak shift or peak splitting occur due to finite
 size effects.}
\label{fig_ene_diff}
\end{figure}

In Fig.\ref{fig_Sq_free}, we show the typical spin structure factors
calculated at the same temperature $T= 0.17|J_1|$ under free BC. Similarly to the case of periodic BC, we find a cubic-symmetric state
(Fig.\ref{fig_Sq_free} (a)) and a non-cubic state of type 1
(Fig.\ref{fig_Sq_free} (b)). By contrast, the non-cubic state of type 2 is never observed in contrast to the case of periodic BC. (In the non-cubic state observed under free BC, the main and sub peaks appear along the direction of $\bm{q} = \frac{2\pi}{a}(h, h+\delta, 0)$ slightly off the high symmetric direction $(h, h,0)$ reflecting the non-cubic character of the ordered states. However, the magnitude of the shift $\delta$ is always of order of $1/L$, apparently vanishing in the thermodynamic limit.)

 We deduce that the type-1 state rather than the type-2 state is a stable quadruple-$q$ state due to the following two reasons. First, the type-1 quadruple-$q$ state appear as a metastable state in both periodic and free BC, while the type-2 quadruple-$q$ state is never realized in free BC. Second, within our mean-field analysis, the type-1-like state with the spin structure factor of unequal peak heights, which corresponds to the $\mathrm{q2}$ state in appendix, has a lower free energy than the type-2-like state with the spin structure factor of equal peak heights, which corresponds to the $\mathrm{q1}$ state in appendix.

In contrast to the mean-field result for $J_2/J_1 \simeq -0.2$, we do not observe  in our MC simulation the double-$q$ metastable state below $T_c$. Since the double-$q$ state can continuously be changed into the quadruple-$q$ state, the double-$q$ state is probably unstable toward the quadruple-$q$ state in this parameter region.

Next, we wish to determine either the cubic sextuple-$q$ state or the
non-cubic quadruple-$q$ state is stable below $T_c$. Since both states
are metastable states surviving for a very long time in the course of MC
simulation, it is extremely difficult to identify which state is
thermodynamically stable by performing fully thermalized simulation
below $T_c$. In order to identify the truly stable state below $T_c$, we
then employ the mixed-phase method \cite{Chern,Creutz}.  In this method, one prepares as an initial state a two-phase coexisting state, where the sextuple-$q$ state or the quadruple-$q$ state occupies each half of the lattice. By thermalizing such a state and by monitoring which ordered state expands during the course of the subsequent simulation, one can determine which phase is more stable.

In Fig.\ref{fig_mix}, we show an example of the MC time evolution of such a two-phase coexisting initial state in our MC simulations. We look at the time evolution of the cubic-symmetry-breaking order parameter $m_c$ for each half of the lattice, defined by
\begin{equation}
m_c \equiv 2\frac{\langle \max \left(\sum_{i} \bm{S}_i\cdot\bm{S}_{i+\delta}\right) \rangle -\langle \min \left(\sum_{i}
 \bm{S}_i\cdot\bm{S}_{i+\delta}\right) \rangle } {\langle \max \left(\sum_{i} \bm{S}_i\cdot\bm{S}_{i+\delta}\right) \rangle +\langle \min \left(\sum_{i}  \bm{S}_i\cdot\bm{S}_{i+\delta}\right) \rangle},
\label{eq_mc}
\end{equation}
where $\max \left(\sum_i\bm{S}_i\cdot\bm{S}_{i+\delta}\right)$ and $\min
\left(\sum_i\bm{S}_i\cdot\bm{S}_{i+\delta}\right)$ represent the maximum
and the minimum values of the sum of the inner product $\bm{S}_i\cdot\bm{S}_{i+\delta}$ among three nearest-neighbor bonds, denoted by $\delta$, emanating from a given site $i$. This quantity measures the extent of the cubic-symmetry breaking in the spin configuration: It tends to be large for a non-cubic quadruple-$q$ state, while it tends to be small for a cubic sextuple-$q$ state. As can be seen from the MC time dependence of $m_c$ shown from Fig.\ref{fig_mix}, at the measuring temperature $T=0.15|J_1|$, the sextuple-$q$ state expands its area, and the system finally settles in the sextuple-$q$ state with smaller $m_c$. By performing such a mixed-phase method for various system size up to $L=24$ for periodic BC, and up to $L=48$ for free BC, we find that the sextuple-$q$ state is stable at least in the temperature range $T \ge 0.15J_1$. Unfortunately, the dynamics of the system becomes so slow with further lowering the temperature that the domain wall between the two ordered states hardly moves, which hampers to determine the stable state at lower temperatures $T<0.15 J_1$.

In the low temperature limit, the entropy contribution to the free energy becomes smaller, while the energy contribution becomes dominant. Thus, the state which has the lower energy should be realized. In Fig.\ref{fig_ene_diff}, we show the energy difference between the sextuple-$q$ state and the quadruple-$q$ state versus the temperature. For both cases of periodic and free BCs, the energy of the quadruple-$q$ state becomes lower than that of the sextuple-$q$ state at low enough temperatures. It means that the quadruple-$q$ state should be realized as a stable state at sufficiently low temperatures.

By combining the results of the energy analysis with those of the mixed-phase method, we conclude the ordering behavior of the $J_1$-$J_2$ model with $J_2/J_1=-0.2$ as follows. With decreasing the temperature,  the system first exhibits a phase transition from the paramagnetic phase to the sextuple-$q$ ordered state at a temperature $T=T_{c1}\simeq 0.178|J_1|$. The transition is of first order. Note that the cubic symmetry of the lattice is still fully respected even below $T_{c1}$ in spite of the existence of the magnetic long-range order characterized by sharp Bragg peaks. With further decreasing the temperature, another phase transition occurs  at $T=T_{c2} < 0.15|J_1|$ into the quadruple-$q$ state. It accompanies the breaking of the cubic symmetry of the lattice. This transition is probably of weakly first order.

Now, we turn to the determination of the detailed spin configurations in the multiple-$q$ states identified above. We begin with the cubic-symmetric sextuple-$q$ state. Guided by our mean-field result and by inspecting the spin configurations obtained by MC simulations, we observe that the ordered spin configuration in the sextuple-$q$ state can be represented by the superposition of six linearly-polarized SDWs as
\begin{align}
 S_x^{(\mu)}(\bm{r}) &= I \Bigl[
 U^\mu_{\bm{q}_1^+}\cos(\bm{q}_1^+\cdot\bm{r}+\theta_1^+) \notag \\
 &\qquad\qquad+U^\mu_{\bm{q}_1^-} \cos(\bm{q}_1^-\cdot\bm{r}+\theta_1^-)\Bigr]
 \notag \\
 S_y^{(\mu)}(\bm{r}) &= I \Bigl[
 U^\mu_{\bm{q}_2^+}\cos(\bm{q}_2^+\cdot\bm{r}+\theta_2^+)\notag \\
 &\qquad\qquad+U^\mu_{\bm{q}_2^-} \cos(\bm{q}_2^-\cdot\bm{r}+\theta_2^-)\Bigr] 
 \notag \\
 S_z^{(\mu)}(\bm{r}) &= I \Bigl[
 U^\mu_{\bm{q}_3^+}\cos(\bm{q}_3^+\cdot\bm{r}+\theta_3^+)\notag\\
 &\qquad\qquad+U^\mu_{\bm{q}_3^-} \cos(\bm{q}_3^-\cdot\bm{r}+\theta_3^-)\Bigr],
\label{model_sextuple}
\end{align}
where $\bm{S}^{(\mu)}(\bm{r}) =
(S_x^{(\mu)}(\bm{r}),S_y^{(\mu)}(\bm{r}),S_z^{(\mu)}(\bm{r}))$ represent
the spin  at the position $\bm{r}$ belonging to sublattice
$\mu$. Any spin configuration generated from the one given by \eqref{model_sextuple} via
global spin rotation in spin space is equally allowed.  The coefficient
$U^\mu_{\bm{q}}$ takes a value unity or $-\alpha$, depending on whether
the site on the sublattice $\mu$ has a NN bond in the direction of the
wavevector $\bm{q}=\bm{q}_i^{\pm}$ ($U^\mu_{\bm{q}_i^{\pm}}=1$) or not
($U^\mu_{\bm{q}_i^{\pm}}=-\alpha$). Indeed, the spin configuration given
by Eq.\eqref{model_sextuple} is fully  consistent with the mean-field result shown in Sec. III and appendix.

In the the case of the quadruple-$q$ state, inspection of the MC data lead us to the ordered-state spin configuration represented by the superposition of two spirals and two linearly-polarized SDWs as
\begin{align}
 S_x^{(\mu)}(\bm{r}) &= I_{xy} \Bigl[
 \tilde{U}^\mu_{\bm{q}_1^+}\cos(\bm{q}_1^+\cdot\bm{r}+\theta_1^+) \notag \\
 &\qquad\qquad+\tilde{U}^\mu_{\bm{q}_1^-} \cos(\bm{q}_1^-\cdot\bm{r}+\theta_1^-)\Bigr]
 \notag \\
 S_y^{(\mu)}(\bm{r}) &= I_{xy} \Bigl[
 \tilde{U}^\mu_{\bm{q}_1^+}\sin(\bm{q}_1^+\cdot\bm{r}+\theta_1^+) \notag \\
 &\qquad\qquad+\tilde{U}^\mu_{\bm{q}_1^-} \sin(\bm{q}_1^-\cdot\bm{r}+\theta_1^-)\Bigr]
 \notag \\
 S_z^{(\mu)}(\bm{r}) &= I_z \Bigl[
 \tilde{U}^\mu_{\bm{q}_2^+}\cos(\bm{q}_2^+\cdot\bm{r}+\theta_2^+)\notag \\
 &\qquad\qquad +\tilde{U}^\mu_{\bm{q}_2^-} \cos(\bm{q}_2^-\cdot\bm{r}+\theta_2^-)\Bigr],
\label{model_quadruple}
\end{align}
where the two wavevectors with stronger intensity in the spin structure factor are associated with $\bm{q}_1^{\pm}$ and those with weaker intensity with $\bm{q}_2^{\pm}$. Any spin configuration generated from the one given by \eqref{model_quadruple} via global spin rotation in spin space is equally allowed. The same rule given above for the sextuple-$q$ state is understood also for the coefficient $\tilde{U}^\mu_{\bm{q}}$. This spin configuration is also fully consistent with the mean-field result of Sec. III and appendix. 

\begin{figure*}
 \includegraphics[width=16cm]{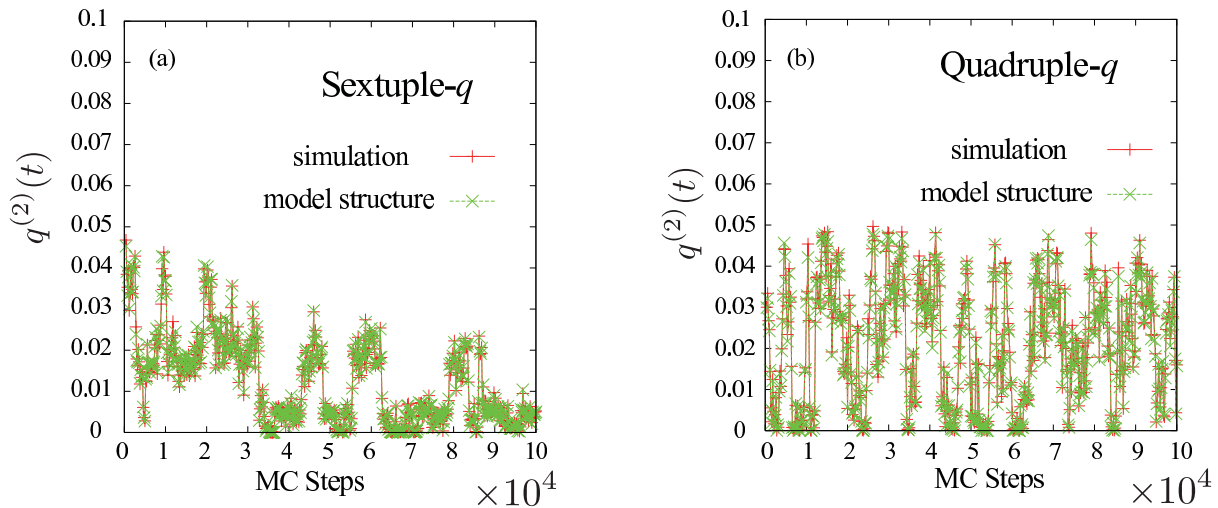}
 \caption{(color online) The Monte-Carlo time dependence of the overlap $q^{(2)}(t)$
 defined by Eq. \eqref{eq_q2_freeze}, calculated at $T=0.17|J_1|$ for a $L=16$
 lattice with periodic boundary conditions, in the cases of (a) the
 sextuple-$q$ state and of (b) the quadruple-$q$ state. The interaction parameter is  $J_2/J_1=-0.2$. Two curves in the figure represent $q^{(2)}(t)$ calculated either  directly from the raw Monte Carlo data or from the proposed model spin
 structure evaluated from the Monte Carlo data: See the text for
 further details. In both cases of (a) and (b), the two curves agree with high precision.}
 \label{fig_freeze}
\end{figure*}

 In order to investigate the nature of fluctuation in the ordered state, we compute the time-dependent spin overlap $q^{(2)}(t)$ defined by
\begin{equation}
 q^{(2)}(t) \equiv \sum_{\alpha,\beta}
 \left[\sum_{i}\sum_{\mu}S^{(\mu)}_{\alpha}(\bm{r}^{(\mu)}_i,t_0)S^{(\mu)}_{\beta}(\bm{r}_i^{(\mu)},t_0+t)\right]^2 ,
 \label{eq_q2_freeze}
\end{equation}
which measures an overlap between the spin configurations at time $t_0$ and at time $t_0+t$.  Note that $q^{(2)}(t)$ is defined so as to be invariant under any global spin rotation in spin space. 

We also employ this quantity to check the validity of the proposed spin structures given by Eqs.\eqref{model_sextuple} and \eqref{model_quadruple}. For this purpose, we compute $q^{(2)}(t)$ in the following two ways: The two spin configurations at time $t_0$ and at subsequent time $t_0+t$ are taken, either (i) from the raw spin configuration data of our equilibrium MC simulations, or (ii) from the  the proposed model spin structure described by Eq.\eqref{model_sextuple} or \eqref{model_quadruple} which is evaluated by Fourier transforming the raw MC data and by extracting the amplitudes ($I$, $I_{xy}$ and $I_z$) and the phases ($\theta_i^\pm$) associated with the critical mode at $q_i^\pm$. In the procedure (ii), the $\alpha$-factor associated with $U$ or $\tilde{U}$ is also taken as a fitting parameter, not a given constant, while the contribution of other non-critical modes are simply neglected. If the proposed model structures properly represent the ordered-state spin configuration, $q^{(2)}(t)$s calculated in the above two ways (i) and (ii) should agree.

\begin{figure*}
 \includegraphics[width=13cm]{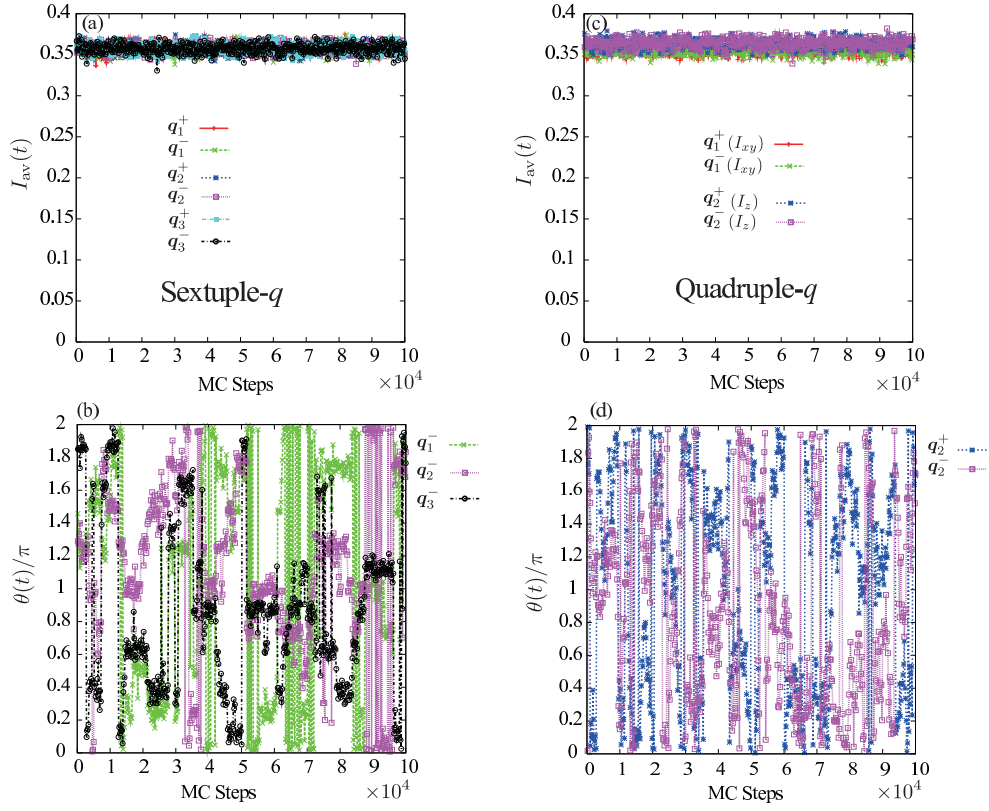}
 \caption{(color online) The Monte-Carlo time dependence of the amplitudes $I$,$I_{xy}$,$I_z$ ((a) and (c)) and the phase factors  $\theta_i^\pm$ ((b) and (d)) for the cases of the sextuple-$q$ state ((a) and (b)) and of the quadruple-$q$ state ((c) and (d)), as obtained by fitting the raw Monte-Carlo data to the model spin structures of Eqs.\eqref{model_sextuple} and \eqref{model_quadruple}. Concerning the amplitudes, plotted here are the average intensities over sublattices $\mu$, {\it i.e.\/}, $I_{\mathrm{av}}=\sqrt{\sum_{\mu}\left[IU^\mu_{\bm{q}}\right]^2/4}$. The temperature is $T=0.17|J_1|$ and the lattice is $L=16$ with periodic boundary conditions. The interaction parameter is  $J_2/J_1=-0.2$. The amplitudes, which contribute to the Bragg intensity, are kept nearly constant in time, whereas the phase factors fluctuate a lot.} 
 \label{fig_int_phase}
\end{figure*}

 In Fig.\ref{fig_freeze}, we show the MC-time dependence of
 $q^{(2)}(t)$ calculated in the two ways (i) and (ii) for the
 sextuple-$q$ state (a), and for the quadruple-$q$ state (b). Indeed, in 
 both cases of the sextuple-$q$ and the quadruple-$q$ states,
 $q^{(2)}(t)$ calculated in the two ways agree with high precision,
 indicating that our model spin structures \eqref{model_sextuple} and
 \eqref{model_quadruple} properly represent the actual ordered-state
 spin configurations. Second, the computed $q^{(2)}(t)$ exhibit a
 significant time variation indicating that the spin configuration
 changes considerably in the course of our MC simulation. Indeed, as can be seen from  Fig.\ref{fig_int_phase}, this time
 dependence comes from the time dependence of the phase factor
 $\theta_i^{\pm}$, whereas the amplitude turns out to be nearly time-independent.  

 Establishing the spin configuration of the ordered state, we wish to
 further investigate the nature of spin fluctuations in the ordered
 state. Two kinds of spin fluctuations associated with the identified
 spin structures are observed. The first one is a fluctuation of the
 phase factors $\theta_i^{\pm}$ as mentioned above. While the amplitudes
 $I$, $I_{xy}$, $I_z$, which contribute to the Bragg intensity, turn out
 to be nearly constant in time (see Fig.\ref{fig_int_phase} (a) and (c)),
 the phase factors $\theta_i^{\pm}$ fluctuate a lot even when the system
 exhibits sharp Bragg peaks (see Fig.\ref{fig_int_phase} (b) and (d)). The second type of spin fluctuation is a necessary outcome of the distribution of ordered spin moments in the multiple-$q$ state, where some spins should reduce their frozen moments exhibiting large fluctuations.

\begin{figure}
 \includegraphics[width=8cm]{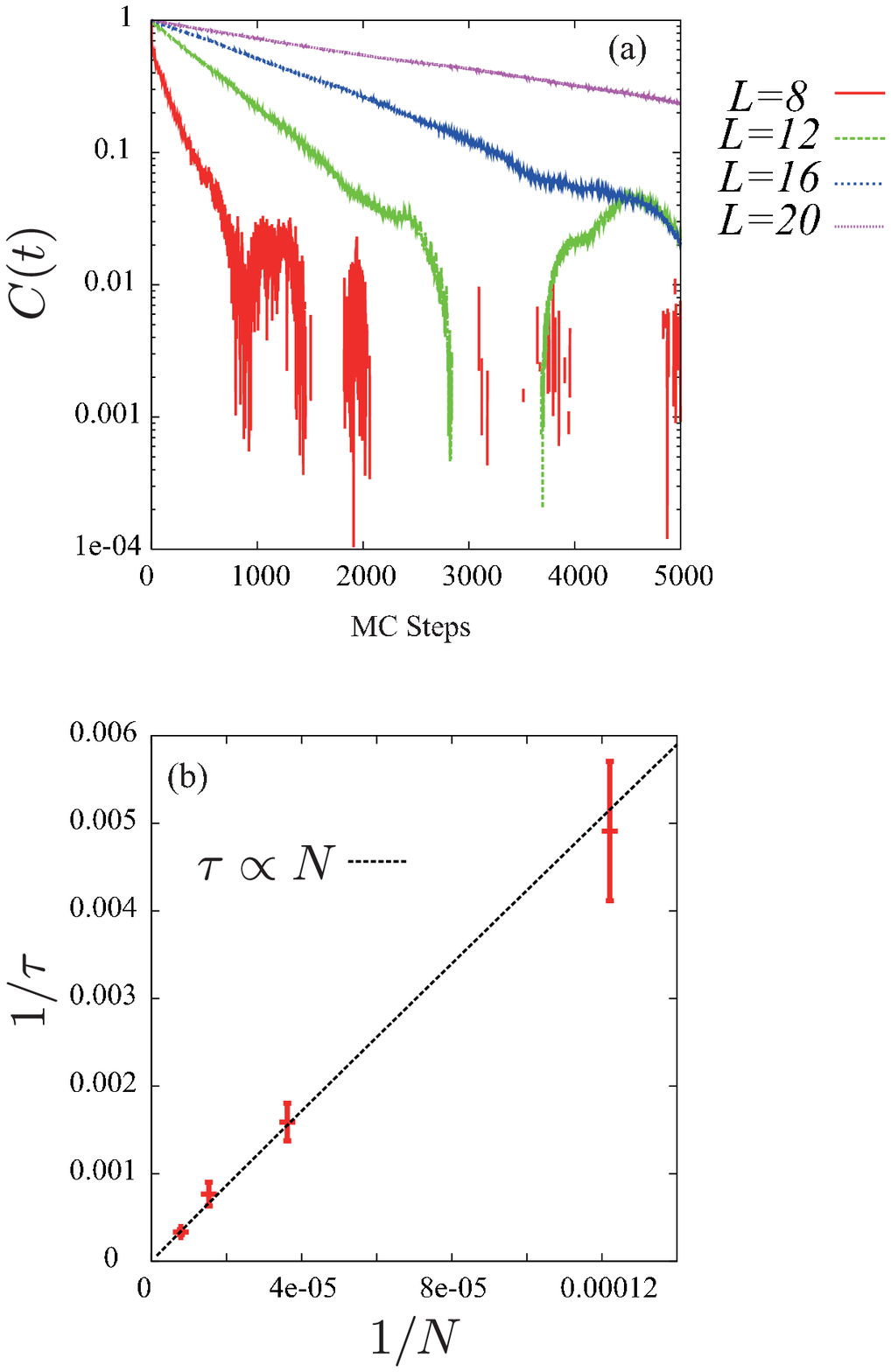}
 \caption{(color online)  (a) The Monte-Carlo time dependence of the
 phase $\theta_1^{\pm}$ autocorrelation function in the quadruple-$q$
 state at $T=0.17J_1$. The lattices are $L=8,12,16$ and $20$ with periodic boundary conditions. (b) The phase autocorrelation time obtained as in (a) is plotted versus the inverse system size $1/N$. Dotted line exhibits a liner fit of the data.}
 \label{fig_phase_auto}
\end{figure}

First, we examine the effect of phase fluctuations. As we have seen in the dynamics of $\theta_i^{\pm}(t)$ shown in Fig.\ref{fig_int_phase} (b) and (d), phase fluctuations do occur in finite-size systems. The question to be addressed is whether such phase fluctuations disappear or not in the thermodynamic limit. For this purpose, we define the phase auto-correlation function by
\begin{equation}
 C(t) \equiv \langle \cos (\theta(t)-\theta(0)) \rangle ,
\end{equation}
where $\theta (t)$ represents $\theta_i^{\pm}(t)$. In Fig.\ref{fig_phase_auto} (a), we show the phase autocorrelation
function of $\theta_1^{\pm}$ in the quadruple-$q$ state. As can be seen
from the figure, the relaxation of the phase becomes slower with
increasing the system size, suggesting that the phase motion might
eventually be locked in the thermodynamic limit. The computed phase
relaxation follows a simple exponential decay characterized by the phase correlation time $\tau$. In Fig.\ref{fig_phase_auto} (b), we plot $1/\tau$ as a function of the inverse system size $1/N$. As can be seen from the figure, the phase relaxation time diverges almost linearly with $N$, and we conclude that the phase motion will eventually be locked in the thermodynamic limit. Hence, spin fluctuations arising from phase fluctuations are finite-size effects and are expected to vanish in the thermodynamic limit.

Next, we discuss the second source of spin fluctuations. Even when phase fluctuations vanish in the thermodynamic limit, spin fluctuations associated with the multiple-$q$ nature of the ordering should still remain. Remember that the multiple-$q$ order entails the the existence of certain fraction of spins with reduced frozen moments, which should be caused by significant local spin fluctuations. 

\begin{figure}
 \includegraphics[width=8cm]{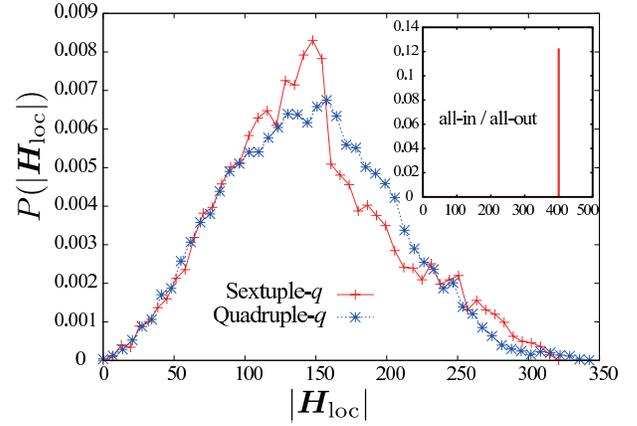}
\caption{(color online) An internal field distribution at an $\mathrm{O}$ site of a pyrochlore oxide $\mathrm{A_2B_2O_6O'}$ with the oxygen-location $x$ parameter $x=0.319$ where $B$ site is magnetic, calculated for an ideal sextuple-$q$ state \eqref{model_sextuple} and quadruple-$q$ state \eqref{model_quadruple}, respectively. Dipolar interactions are assumed as an origin of internal fields. Inset exhibits an internal field distribution  corresponding to a commensurate all-in/all-out structure on the pyrochlore lattice.}
 \label{fig_NMR}
\end{figure}

 Such spin fluctuations intrinsic to the multiple-$q$ order might be observable experimentally via local  probes like NMR. In Fig.\ref{fig_NMR}, we exhibit the distribution of
 internal fields in the sextuple-$q$ (Eqs.\eqref{model_sextuple}) and the
 quadruple-$q$ (Eq.\eqref{model_quadruple}) states probed at an
 $\mathrm{O}$ site of a pyrochlore oxide $\mathrm{A_2B_2O_6O'}$ with the 
 oxygen-location $x$ parameter $x=0.319$ \cite{Pyro_review}. We assume here that the $\mathrm{B}$ site is magnetic, and internal fields are borne by the dipolar interaction. As can be seen from the figure, both the sextuple-$q$ and the quadruple-$q$ states exhibit broad internal-field distributions, in contrast to the one associated with a simple all-in/all-out structure shown in the inset. The latter exhibits a sharp delta-function-like distribution. The observed broad distributions of internal fields mean that the magnitude of frozen spin moments are spatially distributed, as expected from the multiple-$q$ nature of the ordering.

\section{Summary and discussion\label{conclusion}}

 We studied the nature of the multiple-$q$ ordered states realized in the $J_1$-$J_2$ pyrochlore-lattice Heisenberg model by means of a mean-field analysis and a Monte Carlo simulation.

 Performing a mean-field analysis beyond the previous analysis by Reimers {\it et al.}\cite{Reimers1992}, we could explicitly determine the possible multiple-$q$ spin structures of the model, which include a cubic-symmetric sextuple-$q$ state, a non-cubic quadruple-$q$ state and a non-cubic double-$q$ state. We have found that these multiple-$q$ states have considerably lower free energy than that of the single-$q$ spiral, while the free energy difference between these different multiple-$q$ states are rather small. 

With reference to the mean-field results, we also performed an extensive MC simulations of the model, mainly for the case of $J_2/J_1=-0.2$, to determine which state is really stabilized as an ordered phase. As expected from the mean-field analysis, we found that the system exhibited a phase transition from the paramagnetic phase to the multiple-$q$ ordered state characterized by the multiple peaks in the spin structure factor. Recent studies have revealed that such multiple-$q$ states are also stabilized in other frustrated Heisenberg magnets, {\it e.g.} the triangular lattice Heisenberg antiferromagnet with the next-nearest or the third neighbor interactions under magnetic fields \cite{Triangle}. With the help of the mixed-phase method, we found that the cubic-symmetric sextuple-$q$ state is stabilized just blow the transition temperature $T=T_{c1}\simeq 0.178|J_1|$ down to at least $T = 0.15|J_1|$, while, at sufficiently low temperatures, the non-cubic quadruple-$q$ state becomes stable. Hence, another phase transition from the sextuple-$q$ state to the quadruple-$q$ state should occur at a temperature $T =T_{c2} < 0.15 |J_1|$. 

The nature of each transition was also examined. The transition at $T=T_{c1}$ is first order, as was already indicated by previous studies. Note that, at this transition the cubic symmetry of the lattice is still fully preserved. It is remarkable that, in spite of the incommensurate and rather complex nature of the spin order, the cubic symmetry of the lattice is fully respected in the ordered state.  The second transition at $T=T_{c2}$ is weakly first order, and it accompanies a spontaneous breaking of the cubic symmetry of the lattice.

It should be noticed that the phase transition between the sextuple-$q$ and the quadruple-$q$ states analyzed in the present paper is distinct from the transition between the paramagnetic state and the nematic state (or the one between the nematic state and the multiple-$q$ state) as discussed by Chern {\it et al.}\cite{Chern}. We also performed MC simulations for several smaller values of $J_2/|J_1| < 0.1$ where Chern {\it et al.} observed the nematic phase. In such smaller-$J_2/|J_1|$ region, we found  as metastable states both the sextuple-$q$ and the quadruple-$q$ states in the parameter region where Chern {\it et al.} reported the multiple-$q$ state (Chern {\it et al.} did not specify the type of the multiple-$q$ state). Thus, there remains a possibility that, even for smaller $J_2/|J_1|$-values where the nematic phase appears at higher temperature region, another phase transition from the sextuple-$q$ state to the quadruple-$q$ state occurs at a lower temperature.

We also determined the explicit spin configuration of the multiple-$q$ ordered state. The sextuple-$q$ state is a superposition of six SDWs running along the wavevector $(q^\ast,q^\ast,0)$ and its cubic-symmetry counterparts. In this sextuple-$q$ state, the system fully retains a cubic symmetry of the lattice. By contrast,  the quadruple-$q$ state is a superposition of two helices and two SDWs. This state spontaneously breaks the cubic symmetry of the lattice as one can easily see from the fact that two out of six critical wavevectors should vanish in this state. Both the sextuple-$q$ and the quadruple-$q$ spin configurations are consistent with the ones obtained in our mean-field analysis.

In both cases of the sextuple-$q$ and the quadruple-$q$ states, a broad distribution of internal fields was observed. Such broad distributions reflect the distribution of frozen spin moments which arises from the multiple-$q$ nature of the ordering. Hence, if the pyrochlore magnets are in the multiple-$q$ ordered phase we proposed, one should observe two apparently conflicting features, {\it i.e.\/}, the coexistence of sharp Bragg peaks measured by neutron diffraction and of enhanced spin fluctuations measured by local probes such as NMR.  Experimental observation of the multiple-$q$ states as revealed here remains most interesting.

\begin{acknowledgments}
The authors are thankful to Z. Hiroi, S. Maegawa and T. Arima for
useful discussion. This work is supported by Grand-in-Aid for scientific
Research on Priority Areas ``Novel State of Matter Induced by
Frustration'' (19052006). We thank the Supercomputer Center, Institute
for Solid State Physics, University of Tokyo, and the Cyber Media Center,
Osaka University for providing us with the CPU time.
\end{acknowledgments}

\appendix*
\section{details of the mean-field approximation}

In this appendix, we present the details of our mean-field analysis. We consider here twelve critical eigenvectors only, corresponding to six wavevectors of \eqref{eq_qis}. As mentioned in section III, the free-energy difference between different types of multiple-$q$ states arises at the quartic term of the free energy \eqref{eq_f4}. Thus, we calculate here the quartic term of the free energy $f_4$ for several multiple-$q$ structures.

For simplicity, we abbreviate $\bm{\Phi}^{+}_{\bm{q}_i^{\sigma}}$ as
$\bm{\Phi}_{i\sigma}$, where $i=1,2,3$ and $\sigma = \pm$. The quartic
term $f_4$ consists of following four terms, $A_1$, $A_2$, $A_2'$ as and
$A_4$ as 
\begin{multline}
 f_4 =  \frac{9T}{80\left(1+\alpha^2\right)^2} \Bigl[  (2+2\alpha^4) A_1  + 32\alpha^2 A_2 \\+
 8(1+2\alpha^2+\alpha^4) A_2^\prime +64\alpha^2 A_4 \Bigr],
\label{eq_f4_app}
\end{multline}
 where $A_i$ may be regarded to represent the interaction among
$i$ distinct wavevectors, and are given by 
\begin{equation}
 A_1= \sum_{i=1}^{3}\sum_{\sigma=\pm}\left[4\Phi_{i\sigma}^4+2\left|\bm{\Phi}_{i\sigma}\cdot\bm{\Phi}_{i\sigma}\right|^2\right],
\end{equation}
\begin{equation}
A_2=\sum_{i=1}^{3}\left[\Phi_{i+}^2\Phi_{i-}^2
 + \left|\bm{\Phi}_{i+}\cdot\bm{\Phi}_{i-}\right|^2 + \left|\bm{\Phi}_{i+}\cdot\bm{\Phi}_{i-}^\ast\right|^2\right],
\end{equation}
\begin{equation}
A_2'=\sum_{\langle i\neq j \rangle }\sum_{\sigma,\sigma'}\left[\Phi_{i\sigma}^2\Phi_{j\sigma'}^2
 + \left|\bm{\Phi}_{i\sigma}\cdot\bm{\Phi}_{j\sigma'}\right|^2 + \left|\bm{\Phi}_{i\sigma}\cdot\bm{\Phi}_{j\sigma'}^\ast\right|^2\right],
\end{equation}
and
\begin{multline}
A_4 = \sum_{\langle i \neq
 j\rangle}\hspace{-0.6em}\raisebox{0.4em}{$~^\prime$}\mathrm{Re}\Biggl[\left(\bm{\Phi}_{i+}\cdot\bm{\Phi}_{i-}\right)\left(\bm{\Phi}_{j+}^\ast\cdot\bm{\Phi}_{j-}\right)\\
 \quad+\left(\bm{\Phi}_{i+}\cdot\bm{\Phi}_{j+}^\ast\right)\left(\bm{\Phi}_{i-}\cdot\bm{\Phi}_{j-}\right)\\
 +\left(\bm{\Phi}_{i+}\cdot\bm{\Phi}_{j-}\right)\left(\bm{\Phi}_{i-}\cdot\bm{\Phi}_{j+}^\ast\right)\Biggr] ,
\label{A4_eq}
\end{multline}
where $\Phi_{i\sigma}=|\bm{\Phi}_{i\sigma}|$ and $\mathrm{Re}[\cdots]$ means a real part, while $\sum_{\langle i\neq j\rangle}^\prime$ means the summation over
$(i,j)=(1,3), (2,1), (3,2)$. The $A_4$ term represents interactions of
four wavevectors such as $\bm{q}_1^+$, $\bm{q}_1^-$, $\bm{q}_3^+$,$\bm{q}_3^-$, which satisfy $\bm{q}_1^+ +\bm{q}_1^- -\bm{q}_3^+ + \bm{q}_3^-=\bm{0}$.

\subsection{The single-$q$ state}

In the case of the single-$q$ state, only the $A_1$ term contributes to
$f_4$. If we fix the amplitude $\Phi_{i\sigma}^2=m^2$ associated with
the wavevector $\bm{q}^\sigma_{i}$, $f_4$ takes a minimum when
$\bm{\Phi}_{i\sigma}$ satisfy
\begin{equation}
 \bm{\Phi}_{i\sigma}\cdot\bm{\Phi}_{i\sigma} =0.
\end{equation}
This condition means that $\bm{\Phi}_{i\sigma}$ is given by
\begin{equation}
  \bm{\Phi}_{i\sigma} = m e^{i\theta}\left(\bm{e}_1+ i\bm{e}_2\right),
\label{spiral_eq_phi}
\end{equation}
where $\bm{e}_1$ and $\bm{e}_2$ are orthogonal unit vectors in spin
space. In the real space, this $\bm{\Phi}_{i\sigma}$ means a spiral,
\begin{equation}
 \bm{S}(\bm{r}) \propto \cos(\bm{q}^\sigma_{i}\cdot\bm{r} + \theta)\bm{e}_1 -
 \sin(\bm{q}^\sigma_{i}\cdot\bm{r}+\theta)\bm{e}_2.
\end{equation}

By substituting this condition into Eq.\eqref{eq_f4_app}, we get the minimized free energy for the single-$q$ state as
\begin{equation}
 f_4^{\mathrm{(si)}} = \frac{9T}{10\left(1+\alpha^2\right)^2}(1+\alpha^4)m^4.
\end{equation}

\subsection{The double-$q$ state}

In mixing two wavevectors, there are two possible ways: One is to mix a pair of $\bm{q}_i^{+}$ and $\bm{q}_i^{-}$, and the other is to mix $\bm{q}_i^{\pm}$ and $\bm{q}_j^{\pm}$ with $i\neq j$. For $\alpha < 1$, the former always minimizes the free energy, since the coefficients of $A_2$ is smaller than that of $A_2^\prime$. Hence, we consider below a mixture of $\bm{q}_i^{+}$ and $\bm{q}_i^{-}$. In this situation, the $A_2^\prime$-term vanishes identically.

 The $A_2$-term is minimized for
\begin{align}
 \bm{\Phi}_{i+}\cdot\bm{\Phi}_{i-}=0, \notag\\
 \bm{\Phi}_{i+}\cdot\bm{\Phi}_{i-}^\ast=0.
\end{align}
If we assume that $\bm{\Phi}_{i+}$ forms a spiral given by Eq.\eqref{spiral_eq_phi}, these conditions are
satisfied when $\bm{\Phi}_{i-}$ is perpendicular to the spiral plane formed by $\bm{\Phi}_{i+}$, {\it i.e.},
\begin{equation}
 \bm{\Phi}_{i-} \propto e^{i\theta} \bm{e}_3,
\end{equation}
where $\bm{e}_3$ is a unit vector perpendicular to the spiral plane
formed by $\bm{\Phi}_{i+}$,{\it i.e.}, $\bm{e}_3  =  \bm{e}_1\times
\bm{e}_2$.
 In this situation, the spin configuration corresponding to $\bm{\Phi}_{i-}$ forms a spin-density wave (SDW)
\begin{equation}
 \bm{S}(\bm{r}) \propto \cos(\bm{q}^-_{i}\cdot\bm{r} + \theta)\bm{e}_3.
\end{equation}
 Hence, the $A_2$-term favors a mixture of a spiral and a SDW.

As we have seen for the case of the single-$q$ state, $A_1$ is minimized
when $\bm{\Phi}_{i\sigma}$ forms a spiral. The competition between the
$A_1$- and the $A_2$-terms often causes a distortion of a spiral
structure, {\it i.e.},
\begin{equation}
  \bm{\Phi}_{i\sigma} = e^{i\theta_{i\sigma}}\left(\bm{R}_{i\sigma}+ i\bm{I}_{i\sigma}\right),
\label{distspiral_eq_phi}
\end{equation}
where $\bm{R}_{i\sigma}$ and $\bm{I}_{i\sigma}$ are orthogonal real
vectors
\begin{equation}
 \bm{R}_{i\sigma}\perp \bm{I}_{i\sigma},
\end{equation}
while the amplitudes $R_{i\sigma}=|\bm{R}_{i\sigma}|$ and
$I_{i\sigma}=|\bm{I}_{i\sigma}|$ are generally different from each other.
Note that such a distorted (or elliptical) spiral can continuously be transformed to an isotropic spiral
($R_{i\sigma}=I_{i\sigma}$) and a SDW ($R_{i\sigma}=0$ or
$I_{i\sigma}=0$). Thus, we assume both $\bm{\Phi}_{i+}$ and
$\bm{\Phi}_{i-}$ form distorted spirals Eq.\eqref{distspiral_eq_phi},
and minimize $f_4$ with respect to the directions of spiral planes, the
phase factors $\theta_{i\sigma}$, and the amplitudes $R_{i\sigma}$ and $I_{i\sigma}$.

Under this situation, the $A_2$-term is minimized when the two spiral
planes are mutually orthogonal, {\it i.e.},
\begin{equation}
 (\bm{R}_{i+}\times \bm{I}_{i+}) \perp (\bm{R}_{i-}\times \bm{I}_{i-}).
\end{equation}
Then, $\bm{\Phi}_{i+}$ and $\bm{\Phi}_{i-}$ are given by
\begin{align}
  \bm{\Phi}_{i+} &= e^{i\theta_{i+}}\left(R_{i+}\bm{e}_1+i\kappa_{i+}
 I_{i+}\bm{e}_2\right), \notag \\
  \bm{\Phi}_{i-} &= e^{i\theta_{i-}}\left(R_{i-}\bm{e}_3+ i\kappa_{i-} I_{i-}\bm{e}_1\right),
\end{align}
where $\kappa_{i\sigma}=\pm 1$ represents the chirality of each spiral. In this structure, the two spirals share one of the basis vectors $\bm{e}_1$ while the other two basis vectors $\bm{e}_2$ and $\bm{e}_3$ are mutually orthogonal.  

By substituting $\bm{\Phi}_{i\sigma}$ into Eq.\eqref{eq_f4_app}, $f_4$ is obtained as
\begin{widetext}
 \begin{equation}
  f_4 = \frac{9T}{80\left(1+\alpha^2\right)^2} \Biggl\{2(1+\alpha^2)
 \left\{4\left(\Phi_{i+}^4+\Phi_{i-}^4\right) +
 2\left[\left(R_{i+}^2-I_{i+}^2\right)^2 + \left(R_{i-}^2-I_{i-}^2\right)^2\right]\right\}
 +32\alpha^2\left[\Phi_{i+}^2\Phi_{i-}^2 +2R_{i+}^2I_{i-}^2\right]\Biggr\}.
 \end{equation}
\end{widetext} 
Note that this $f_4$ is independent of the phase factor
$\theta_{i\sigma}$ and the chirality $\kappa_{i\sigma}$.
By minimizing $f_4$ with keeping $\Phi_{i+}^2+\Phi_{i-}^2=m^2$ fixed, we find that there are two types of double-$q$
structure, {\it i.e.\/}, a mixture of two spirals (the $\mathrm{d1}$
state) and a mixture of a spiral and a SDW (the $\mathrm{d2}$ state). 

In the case of a mixture of two spirals (the $\mathrm{d1}$ state), $f_4$
takes an extremum
\begin{equation}
 f^{(\mathrm{d1})}=\frac{9T}{40\left(1+\alpha^2\right)^2}
  \left[3+4\alpha^2+3\alpha^4 -\frac{\left(1+\alpha^4\right)^2}{\left(1+\alpha^2\right)^2} \right]m^4,
\label{eq_fd1}
\end{equation}
for 
\begin{align}
 R_{i+}&=I_{i-}, \notag \\
 I_{i+}&=R_{i-}, \notag \\
 R^2_{i+} &= \frac{1+\alpha^4}{4\left(1+\alpha^2\right)^2}m^2,\notag\\
 I^2_{i+} &= \frac{1+4\alpha^2+\alpha^4}{4\left(1+\alpha^2\right)^2}m^2.
\end{align}
In the $\mathrm{d1}$ state, two spirals are distorted ($R_{i+} <
I_{i+}$, $I_{i-} <R_{i-}$) to lower the $A_2$-term. For $\alpha^2 <
\alpha_c^2=(2-\sqrt{3})$ ($\alpha_c \simeq 0.518$), this $\mathrm{d1}$ state
becomes a minimum of $f_4$, while for $\alpha > \alpha_c$ it becomes locally unstable.

In the case of a mixture of a spiral with $\bm{q}^+_{i}$ and a SDW with
$\bm{q}^-_{i}$ (the $\mathrm{d2}$ state), $f_4$ takes an extremum when the spiral
is isotropic and the SDW is orthogonal to the spiral as
\begin{align}
 R_{i+}&=I_{i+},\notag \\
 I_{i-}&= 0,\notag\\
 R^2_{i+}&=\frac{3-4\alpha^2+3\alpha^4}{2\left(5-8\alpha^2+5\alpha^4\right)}m^2,\notag\\
 R^2_{i-}&=\frac{2-4\alpha^2+2\alpha^4}{5-8\alpha^2+5\alpha^4}m^2,
\end{align}
where $f_4$ is calculated to be
\begin{multline}
f_4^{(\mathrm{d2})}=\frac{9T}{20\left(1+\alpha^2\right)^2}
 \Bigl[-\frac{(3-4\alpha^2+3\alpha^4)^2}{5-8\alpha^2+5\alpha^4}\\ +3\left(1+\alpha^4\right)\Bigr]m^4.
\label{f4_d2}
\end{multline}
For $\alpha^2 > \alpha^2_c= (2-\sqrt{3})$ ($\alpha_c \simeq 0.518$), this $\mathrm{d2}$ state
becomes a minimum of $f_4$, while for $\alpha < \alpha_c$ it becomes locally unstable. 

\begin{figure}
 \includegraphics[width=7cm]{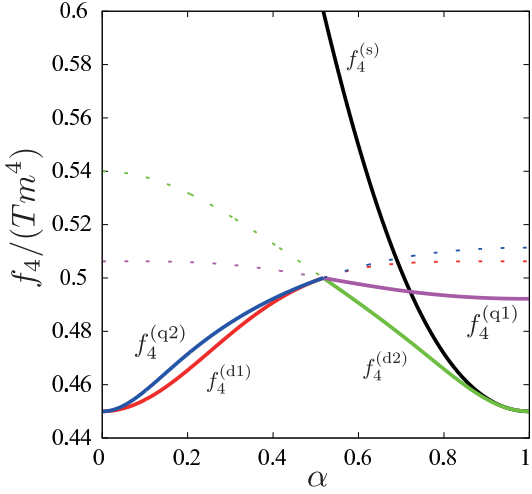}
 \caption{(color online) Quartic terms of the free energy $f_4$ of the
 single-$q$ spiral state (the $s$ state), of the two kinds of double-$q$ states
 (the $\mathrm{d1}$ and $\mathrm{d2}$ states) and of the two kinds of
 quadruple-$q$ states (the $\mathrm{q1}$ and $\mathrm{q2}$ states) are
 plotted versus the parameter $\alpha$. For $\alpha \gtrsim 0.518$, the
 $\mathrm{d1}$ state and $\mathrm{q2}$ state become
 unstable, while for $\alpha \lesssim 0.518$ the $\mathrm{d2}$ and the
 $\mathrm{q1}$ states become unstable. The unstable regions are
 indicated by the thin dotted lines. 
}
 \label{fig_FreeEnergy_ap}
\end{figure}

In Fig.\ref{fig_FreeEnergy_ap}, we plot $f_4/T$ of the single-$q$
 spiral state 
 (the $\mathrm{s}$ state) and of the two kinds of double-$q$ states (the
 $\mathrm{d1}$ and $\mathrm{d2}$ states) as a function of $\alpha$. Note
 that the critical value $\alpha_c$ 
 ($\alpha_c^2=2-\sqrt{3}$) of the stability is in common between the
 $\mathrm{d1}$ state and the $\mathrm{d2}$ state. 
Thus, the mixture of two spirals (the $\mathrm{d1}$
 state) is stable for $\alpha < \alpha_c\simeq 0.518$, while the mixture
 of a spiral and a SDW (the $\mathrm{d2}$ state) is stable for $\alpha >
 \alpha_c$. Since $\alpha$ is in the range of $\alpha \simeq 0.4$ for
 the realistic parameter value of $J_2/J_1$, the mean-field
 approximation favors the $\mathrm{d1}$ state over the $\mathrm{d2}$ state in the $J_1$-$J_2$ model.

\subsection{The quadruple-$q$ state}

When four wavevectors coexist, the $A_4$-term contributes to the free energy. Since the increase of the number of mixed wavevectors tends to enhance the contribution of the $A_2$- and the $A_2^\prime$-terms, it is necessary to lower the $A_4$-term to stabilize the quadruple-$q$ state.  As an example, we consider here a mixture of $\bm{q}_1^{\pm}$  and $\bm{q}_3^{\pm}$.

Since spiral structures tend to lower the $A_1$-term while SDW
structures tend to lower the $A_2$- and the $A_2^\prime$-terms, we again
assume that four $\bm{\Phi}_{i\sigma}$s form distorted spirals described
by Eq.\eqref{distspiral_eq_phi}, and minimize $f_4$ with respect to the
directions of spiral planes, the phase factors $\theta_{i\sigma}$, and the amplitudes $R_{i\sigma}$ and $I_{i\sigma}$.

First, we optimize the spiral plane. In order to lower the $A_2$- and the $A_2^\prime$-terms, the ideal situation would be that
four spiral planes are mutually orthogonal. However, this is simply
impossible for the Heisenberg spin. Since the coefficient of $A_2$ is
smaller than that of $A_2^\prime$, next way might be that a pair of
$\bm{q}_1^{\pm}$ (or $\bm{q}_3^{\pm}$) shares a common spiral axis which
is orthogonal to the others like, 
\begin{align}
 (\bm{R}_{1+}\times \bm{I}_{1+}) &\parallel (\bm{R}_{1-}\times
 \bm{I}_{1-}), \notag \\
 (\bm{R}_{1+}\times \bm{I}_{1+}) &\perp (\bm{R}_{3+}\times
 \bm{I}_{3+}), \notag \\
 (\bm{R}_{1+}\times \bm{I}_{1+}) &\perp (\bm{R}_{3-}\times
 \bm{I}_{3-}).
\end{align}
 Remaining spiral planes of $\bm{q}_3^{\pm}$ prefer to be orthogonal to each other if we consider only the $A_2$-term. However, if we consider also the $A_4$-term, the sum of the $A_2$- and the $A_4$-terms takes a minimum when the pair of $\bm{q}_3^{\pm}$ also shares a common spiral plane orthogonal to that of $\bm{q}_1^{\pm}$ like
\begin{equation}
 (\bm{R}_{3+}\times \bm{I}_{3+}) \parallel (\bm{R}_{3-}\times
 \bm{I}_{3-}),
\end{equation}
as long as they are not reduced to the double-$q$ or the single-$q$ states. 
Thus, four $\bm{\Phi}_{i\sigma}$s are given by
\begin{align}
  \bm{\Phi}_{1+} &= e^{i\theta_{1+}}\left(R_{1+}\bm{e}_1+i\kappa_{1+}
 I_{1+}\bm{e}_2\right), \notag \\
  \bm{\Phi}_{1-} &= e^{i\theta_{1-}}\left(R_{1-}\bm{e}_1+i\kappa_{1-} I_{1-}\bm{e}_2\right),\notag\\
  \bm{\Phi}_{3+} &= e^{i\theta_{3+}}\left(R_{3+}\bm{e}_3+i\kappa_{3+}
 I_{3+}\bm{e}_1\right), \notag \\
  \bm{\Phi}_{3-} &= e^{i\theta_{3-}}\left(R_{3-}\bm{e}_3+i\kappa_{3-} I_{3-}\bm{e}_1\right),
\label{q4_phis}
\end{align}
where $\kappa_{i\sigma}=\pm 1$ represents the chirality of the spiral.

Next, we optimize the phase factor, $\theta_{i\sigma}$,
and the chirality, $\kappa_{i\sigma}$. Although the $A_1$-, the $A_2$-, and
the $A_2'$-terms are independent of the phase factor and the chirality, the
$A_4$-term depends on them. By substituting Eq.\eqref{q4_phis} into
Eq.\eqref{A4_eq}, the $A_4$-term is calculated as  
 \begin{multline}
  A_4 =
  \cos\left(\theta_{1+}+\theta_{1-}-\theta_{3+}+\theta_{3-}\right) \\
  \hspace{-4em}\times
  \Bigl[\left(R_{1+}R_{1-}-\kappa_{1+}\kappa_{1-}I_{1+}I_{1-}\right) \\ 
 \times \left(R_{3+}R_{3-}+\kappa_{3+}\kappa_{3-}I_{3+}I_{3-}\right) \\
  + 2\kappa_{3+}\kappa_{3-}R_{1+}R_{1-}I_{3+}I_{3-}\Bigr].
 \end{multline}
Then, the $A_4$-term is minimized for
\begin{align}
 \cos\left(\theta_{1+}+\theta_{1-}-\theta_{3+}+\theta_{3-}\right)&=-1,
 \notag\\
 \kappa_{1+}\kappa_{1-}&=-1,\notag\\
 \kappa_{3+}\kappa_{3-}&=1,
 \label{phase_conditions}
\end{align}
where $A_4$ is given by
 \begin{multline}
  A_4 = -
   \Bigl[\left(R_{1+}R_{1-}+I_{1+}I_{1-}\right)\left(R_{3+}R_{3-}+I_{3+}I_{3-}\right)
  \\
  + 2R_{1+}R_{1-}I_{3+}I_{3-}\Bigr].
 \end{multline}

By substituting Eqs.\eqref{q4_phis} and \eqref{phase_conditions} into
Eq.\eqref{eq_f4_app}, the optimized $f_4$ is given by
\begin{widetext}
\begin{multline}
 f_4 = \frac{9T}{80\left(1+\alpha^2\right)^2} \Biggl\{2(1+\alpha^4)\sum_{i=1,3}\sum_{\sigma=\pm}
 \left[4\Phi_{i\sigma}^4+2\left(R_{i\sigma}^2-I_{i\sigma}^2\right)^2\right]
 +32\alpha^2 \sum_{i=1,3}\left[\Phi_{i+}^2\Phi_{i-}^2+2\left(R_{i+}^2R_{i-}^2 + I_{i+}^2I_{i-}^2\right)\right]\\
 + 8 (1+2\alpha^2+\alpha^4) \sum_{\sigma=\pm}\sum_{\sigma'=\pm}\left[\Phi_{1\sigma}^2\Phi_{3\sigma'}^2 +2R_{1\sigma}^2I_{3\sigma'}^2\right]\\
-64\alpha^2    \Bigl[\left(R_{1+}R_{1-}+I_{1+}I_{1-}\right)\left(R_{3+}R_{3-}+I_{3+}I_{3-}\right)
  + 2R_{1+}R_{1-}I_{3+}I_{3-}\Bigr] \Biggr\}.
\end{multline}
\end{widetext}
By minimizing $f_4$ for a fixed $\Phi_{1+}^2+\Phi_{1-}^2+\Phi_{3+}^2+\Phi_{3-}^2=m^2$, we find that there are two types of quadruple-$q$
structure, {\it i.e.\/}, a mixture of four distorted spirals (the $\mathrm{q1}$
state) and a mixture of two spirals and two SDWs (the $\mathrm{q2}$ state). 

In the case of a mixture of four spirals (the $\mathrm{q1}$ state), $f_4$ takes an extremum,
\begin{multline}
 f_4^{(\mathrm{q1})} =
  \frac{9T}{80\left(1+\alpha^2\right)^2}\Biggl[5+4\alpha^2+5\alpha^4 \\ 
\hspace{10em}-\frac{1-8\alpha^2-14\alpha^4-8\alpha^6+\alpha^8}{2\left(1+\alpha^2\right)^2}\Biggr]m^4,\\ 
\label{eq_fq1}
\end{multline}
with
\begin{align}
 R_{1+}&=R_{1-}=I_{3+}=I_{3-}, \notag \\
 I_{1+}&=I_{1-}=R_{3+}=R_{3-}, \notag \\
 R^2_{1+} &= \frac{1+4\alpha^2+\alpha^4}{16\left(1+\alpha^2\right)^2}m^2,\notag\\
 I^2_{1+} &= \frac{3+4\alpha^2+3\alpha^4}{16\left(1+\alpha^2\right)^2}m^2.
\end{align}
In this $\mathrm{q1}$ state, similarly to the $\mathrm{d1}$ state, the four spirals are distorted
($R_{1\sigma} < I_{1\sigma}$, $I_{3\sigma} < R_{3\sigma}$) to lower the $A_2'$-term. For $\alpha^2 >
\alpha_c^2=(2-\sqrt{3})$ ($\alpha_c \simeq 0.518$), this $\mathrm{q1}$ state
becomes a minimum of $f_4$, while for $\alpha < \alpha_c$ it becomes locally unstable.

Next, we consider a mixture of two spirals and two SDWs (the
$\mathrm{q2}$ state). In the $\mathrm{q2}$ state, a set of
$\bm{\Phi}_{i+}$ and $\bm{\Phi}_{i-}$ ($i=1$ or $3$) form spirals, while
the other set of $\bm{\Phi}_{j+}$
and $\bm{\Phi}_{j-}$ ($j\neq i$) form SDWs perpendicular the spiral
plane. As an example, we choose here a mixture of two spirals
characterized by $\bm{q}_{1\pm}$ and two SDWs characterized by  $\bm{q}_{3\pm}$. In
this situation, $f_4$ takes an extremum when two spirals are isotropic,
\begin{align}
R_{1+}&=I_{1+},\notag\\
R_{1-}&=I_{1-},
\end{align}
and two SDWs are perpendicular to the spiral plane,
\begin{align}
I_{3+}&=0,\notag\\
I_{3-}&=0.
\end{align}
In addition, their amplitudes satisfy
\begin{align}
 R^2_{1+}=R^2_{1-}&= \frac{1+12\alpha^2+\alpha^4}{4\left(1+20\alpha^2+\alpha^4\right)}m^2,\notag\\
 R^2_{3+}=R^2_{3-}&= \frac{4\alpha^2}{1+20\alpha^2+\alpha^4}m^2.
\end{align}
The optimized $f_4$ is then calculated to be
\begin{multline}
 f_4^{(\mathrm{q2})} =
 \frac{9T}{40\left(1+\alpha^2\right)^2}\Biggl[-\frac{\left(1+12\alpha^2+\alpha^4\right)^2}{1+20\alpha^2+\alpha^4}\\
                                       +3\left(1+4\alpha^2+\alpha^4\right)\Biggr]m^4.
\label{eq_fq2}
\end{multline}
For $\alpha^2 < \alpha_c^2=(2-\sqrt{3})$ ($\alpha_c \simeq 0.518$), this $\mathrm{q2}$ state becomes a minimum of $f_4$, while for $\alpha > \alpha_c$ it becomes locally unstable.

Thus, for the quadruple-$q$ state, we find two metastable structures,
$\mathrm{q1}$ and $\mathrm{q2}$. In Fig.\ref{fig_FreeEnergy_ap}, we plot
$f_4/T$ of these two metastable quadruple-$q$ structures, given by
\eqref{eq_fq1} and \eqref{eq_fq2}, respectively, as a function of
$\alpha$. Note that the critical value $\alpha_c\simeq 0.518$
 ($\alpha_c^2=2-\sqrt{3}$) of the stability is common between the
 $\mathrm{q1}$ state and the $\mathrm{q2}$ state. Since $\alpha$ is in the range of $\alpha \sim 0.4$ for the realistic parameter value of $J_2/J_1$, the structure characterized by a mixture of two
spirals and two SDWs (the $\mathrm{q2}$ state) is the stable one
between the two metastable quadruple-$q$ structures.

\subsection{The sextuple-$q$ state}

We finally discuss the case where all six wavevectors coexist, {\it i.e.}, the sextuple-$q$ state. In this situation, although almost all types of combinations of spirals and SDWs turn out to be unstable, a mixture of six SDWs can be locally stable.

For a mixture of six SDWs, $f_4$ is minimized when SDWs of
$\bm{q}_{i}^+$ and $\bm{q}_{i}^-$ run along a common axis while SDWs of $\bm{q}_{i}^\pm$ and
$\bm{q}_{j}^\pm$ ($i\neq j$) are mutually orthogonal, {\it i.e.},
\begin{equation}
 \bm{\Phi}_{i\pm} = \Phi_{i\pm}e^{i\theta_{i\pm}}\bm{e}_i, \qquad (i=1,2,3).
 \label{eq_phi_q6}
\end{equation}
As in the case of the quadruple-$q$ state, the phase
factors are optimized to minimize the $A_4$-term. By substituting
Eq.\eqref{eq_phi_q6} into Eq.\eqref{A4_eq}, the $A_4$-term is calculated
as 
 \begin{multline}
  A_4 =
  \cos\left(\theta_{1+}+\theta_{1-}-\theta_{3+}+\theta_{3-}\right)
  \Phi_{1+}\Phi_{1-}\Phi_{3+}\Phi_{3-} \\
  \hspace{2em}+\cos\left(\theta_{2-}+\theta_{2+}-\theta_{1+}+\theta_{1-}\right)
  \Phi_{2+}\Phi_{2-}\Phi_{1+}\Phi_{1-} \\
  \hspace{4em}+\cos\left(\theta_{3-}+\theta_{3+}-\theta_{2+}+\theta_{2-}\right)
  \Phi_{3+}\Phi_{3-}\Phi_{2+}\Phi_{2-}.
 \end{multline}
Then, the $A_4$-term is minimized for
\begin{align}
 \cos\left(\theta_{1+}+\theta_{1-}-\theta_{3+}+\theta_{3-}\right)&=-1,\notag\\
 \cos\left(\theta_{2+}+\theta_{2-}-\theta_{1+}+\theta_{1-}\right)&=-1,\notag\\
 \cos\left(\theta_{3+}+\theta_{3-}-\theta_{2+}+\theta_{2-}\right)&=-1,
\label{phase_conditions_q6}
\end{align} 
where $A_4$ is given by
\begin{multline}
 A_4= -   \Phi_{1+}\Phi_{1-}\Phi_{3+}\Phi_{3-} -
  \Phi_{2+}\Phi_{2-}\Phi_{1+}\Phi_{1-} \\
 -  \Phi_{3+}\Phi_{3-}\Phi_{2+}\Phi_{2-}.
\end{multline}
From Eqs.\eqref{eq_phi_q6} and \eqref{phase_conditions_q6}, the optimized
$f_4$ is given by
\begin{widetext}
\begin{multline}
 f_4=\frac{9T}{80(1+\alpha^2)^2}\Biggl[12(1+\alpha^4)\sum_i\left(\Phi_{i+}^4+\Phi_{i-}^4\right)
 +96\alpha^2 \sum_i\Phi_{i+}^2\Phi_{i-}^2
 +8\left(1+2\alpha^2+\alpha^4\right)\sum_{\langle i,j
 \rangle}\sum_{\sigma,\sigma'=+,-}\Phi_{i\sigma}^2\Phi_{j\sigma'}^2 \\
 -64\alpha^2 \sum_{\langle i,j
 \rangle} \Phi_{i+}\Phi_{i-}\Phi_{j+}\Phi_{j-}\Biggr].
\end{multline}
\end{widetext}
If we fix the total amplitude $m^2=\sum_{i}\sum_{\sigma}\Phi_{i\sigma}^2$, $f_4$ is minimized when the amplitudes of six SDWs are in common, {\it i.e.},
\begin{equation}
 \Phi_{i\sigma}^2=\frac{m^2}{6}.
\end{equation}
In this situation, $f_4$ of the sextuple-$q$ state is reduced to
\begin{equation}
 f_4^{\mathrm{(se)}}=\frac{3T}{40(1+\alpha^2)^2}\left(7+12\alpha^2+7\alpha^4\right)m^4.
\end{equation}

\end{document}